\documentclass[universe,article,accept,pdftex,moreauthors]{Definitions/mdpi} 
\firstpage{1} 
\pubvolume{1}
\issuenum{1}
\articlenumber{0}
\pubyear{2025}
\copyrightyear{2025}
\externaleditor{Firstname Lastname} 
\datereceived{ } 
\daterevised{ } 
\dateaccepted{ } 
\datepublished{ } 
\hreflink{https://doi.org/} 

\Title{Super-Accreting Active Galactic Nuclei as Neutrino Sources}

\TitleCitation{Super-Accreting Active Galactic Nuclei as Neutrino Sources}

\Author{Gustavo E. Romero 
 $^{1, 2,}$*\orcidA{} and Pablo Sotomayor $^{2}$\orcidB{}}

\AuthorNames{Gustavo Romero and Pablo Sotomayor}

\isAPAStyle{%
       \AuthorCitation{Romero, G.E., \& Sotomayor, P.}
         }{%
        \isChicagoStyle{%
        \AuthorCitation{Gustavo E. Romero, and Pablo Sotomayor.}
        }{
        \AuthorCitation{Romero, G.E.; Sotomayor, P.}
        }
}

\address{%
$^{1}$ \quad Instituto Argentino de Radioastronom\'{\i}a ( 
CCT-La Plata; 
 CONICET,
 CICPBA, UNLP), C.C. 
 No. 5, \linebreak  1894 Villa Elisa, Argentina\\
$^{2}$ \quad Facultad de Ciencias Astron\'omicas y Geof\'{\i}sicas, Universidad Nacional de La Plata, Paseo del Bosque s/n, 1900 La Plata, Argentina; pablosotomayor@fcaglp.unlp.edu.ar
}

\corres{Correspondence:  romero@iar.unlp.edu.ar}

\abstract{Active galactic nuclei (AGNs) often exhibit broad-line regions (BLRs), populated by high-velocity clouds in approximately Keplerian orbits around the central supermassive black hole (SMBH) at subparsec scales. During episodes of intense accretion at super-Eddington rates, the accretion disk can launch a powerful, radiation-driven wind. This wind may overtake the BLR clouds, forming bowshocks around them. Two strong shocks arise: one propagating into the wind, and the other into the cloud. If the shocks are adiabatic, electrons and protons can be efficiently accelerated via a Fermi-type mechanism to relativistic energies. In sufficiently dense winds, the resulting high-energy photons are absorbed and reprocessed within the photosphere, while neutrinos produced in inelastic
 $pp$ collisions escape.
In this paper, we explore the potential of super-accreting AGNs as neutrino sources. We propose a new class of neutrino emitter: an AGN lacking jets and gamma-ray counterparts, but hosting a strong, opaque, disk-driven wind. As a case study, we consider a supermassive black hole with $M_{\rm BH} = 10^{6}\,M_{\odot}$ and accretion rates consistent with tidal disruption events (TDEs). We compute the relevant cooling processes for the relativistic particles under such conditions and show that super-Eddington accreting SMBHs can produce detectable neutrino fluxes with only weak electromagnetic counterparts. The neutrino flux may be observable by the next-generation IceCube Observatory (IceCube-Gen2) in nearby galaxies with a high BLR cloud filling factor. For galaxies hosting more massive black holes, detection is also possible with moderate filling factors if the source is sufficiently close, or at larger distances if the filling factor is high.
Our model thus provides a new and plausible scenario for high-energy extragalactic neutrino sources, where both the flux and timescale of the emission are determined by the number of clouds orbiting the black hole and the duration of the super-accreting phase.}

\keyword{galaxies: 
 active; galaxies: nuclei; galaxies: Seyfert;  ISM: jets and 
 outflows; radiation mechanisms: non-thermal} 

\begin{document}

\section{Introduction}

Accretion onto black holes is the most efficient mechanism known for transforming gravitational energy into radiation. Angular momentum of the accretion flow drives the formation of a disk around the black hole. This phenomenon is often accompanied by the ejection of outflows and jets, that can strongly affect the environment. The main parameter that determines the physical properties of the accretion disk and the ejection of outflows is the ratio between the actual and the critical accretion rates. The critical rate is defined as $\dot{M}_{\rm cr} = 4{\pi}GMm_{\rm p}/c\sigma_{\rm es}$, where $\sigma_{\rm es}$ is the cross section for Thompson scattering by free electrons. When the rate is super-Eddington (i.e., $\dot{M} \gg \dot{M}_{\rm cr}$) the disk becomes geometrically and optically thick, and it is supported by radiation pressure \citep{paczynsky1980, abramowicz1988, wiita1982}. In these systems, the bolometric luminosity is lower than the accreted power because in the inner region of the disk the photons are trapped and advected towards the black hole. This process can drastically reduce the emerging radiation flux; see  
 \citep{ohsuga2005} and references therein. 
 Semi-analytical models often assume that the accretion rate in the disk is regulated to $\dot{M}_{\rm cr}$ and the excess of matter is ejected as a powerful wind driven by radiation pressure see e.g., ref.~\citep{fukue2004}.\par

Two dimensional, long-term radiation-hydrodynamic simulations confirm the main features of steady supercritical accretion onto stellar black holes inferred with the semi-analytical models \citep{ohsuga2007}. \citet{jiang2019} implemented three-dimensional radiation-magnetohydrodynamic simulations of super-Eddington accretion disks onto SMBHs and verified the possibility of fast outflows driven by the intense radiation pressure of the disk. Recently, \citet{ashanina2022} investigated super-Eddington accretion onto both stellar and SMBHs and analyzed the results obtained for different closure schemes. They found that the intensity of the radiation field must be calculated accurately since approximate closure schemes can give unphysical results for the radiation flux on the axis of rotation of the disk. However, there are no remarkable differences in the physical properties of the disk winds when a closure scheme is used as in some previous works \citep{ohsuga2011,McKinney2014,sadowski2014}. Therefore, semi-analytical investigations and sophisticated numerical simulations are essentially in agreement about that radiation drives fast optically thick winds from the accretion disks in the super-Eddington regime. This is also supported by observational evidence from stellar compact objects \citep{zhou2019}, and SMBHs \citep{liu2019,berdina2021}.\par

When the clouds that form the BLR are overtaken by the disk wind, bowshocks form around the clouds as the collision is highly supersonic. Investigating these interactions is essentially dealing with the problem of a  mildly relativistic fluid in the presence of moving obstacles. In each wind-cloud interaction, two shock waves are produced: a shock in the wind, and a shock in the cloud. If any of these shocks is adiabatic, ions and electrons will be efficiently accelerated up to relativistic velocities by Fermi-type particle acceleration. The electrons will emit synchrotron radiation because the interaction with environmental magnetic fields. This provides a mechanism that contributes to nonthermal radio emission for radio-quiet AGNs, objects for which the origin of the core radio emission is unclear; see~\citep{panessa2019} for an updated review. 
\par

Previously, \citet{sotomayor2022} modeled the nonthermal radiation in super-accreting AGNs originated in the interaction between the disk wind and the BLR clouds. Given the velocities of the wind and the clouds, the nature of the shocks depends on the ratio of densities $n_{\rm w}/n_{\rm c}$ in the region of interaction, where $n_{\rm w}$ and $n_{\rm c}$ are the wind and cloud densities, respectively. 
For the broadband radiation to escape to the observer, the interactions must occur outside the photosphere of the wind. This condition implies that the wind density in the interaction region should be much lower than the density of the clouds. In addition, the wind velocity is a few times higher than the velocity of the clouds. Therefore, the shocks in the wind are typically adiabatic and the shocks in the clouds result~radiative.\par 

Another possible scenario not explored in \cite{sotomayor2022} is when the wind-cloud interactions occur well below the wind photosphere, so shocks in the clouds are adiabatic whereas shocks in the wind become radiative. This situation occurs in AGNs with compact BLRs and with very high accretion rates (and hence very dense winds). The problem then is similar to that investigated by \citet{delValle2018} and \citet{delValle2022MNRAS} for adiabatic shocks in high-velocity clouds colliding with the galactic disk. In the extragalactic scenario, the BLR clouds become acceleration sites for both relativistic electrons and protons. The gamma rays created in $pp$ inelastic collisions within the cloud can be absorbed by the wind, but neutrinos resulting in the same interactions can freely escape.\par
Interest in non-blazar neutrino sources has been reignited by IceCube reports of hot spots likely associated with nearby Seyfert galaxies, such as NGC 1068 \cite{IceCube2022}. Several theoretical models have been proposed invoking a variety of mechanisms e.g.,~\citep{alvarez2004,inoue2020,inoue2021,murase2020,Murase2022ApJ,murase2022,kheirandish2021,anchordoqui2022,kurahashi2022}.
Here we explore a different scenario that yields mostly orphan neutrino sources from the cores of Seyfert-like nuclei, where production is based on $pp$ interactions. Throughout the paper we use tidal disruption events as our benchmark population for the general calculations; this choice is motivated by the observed TDE demographics. Specifically, we aim at estimating the neutrino output from wind-cloud interactions in super-accreting 
 AGNs\endnote{After completing our work, we became aware of a paper by \citet{huang2024}, which applied a similar approach to the case of NGC 1068. Our goal here is to develop a more general and detailed model.}. We shall assume typical parameters for relativistic particle injection through diffusive shock acceleration (DSA) in adiabatic shocks, and we shall calculate the resulting steady state particle spectrum. We focus on lepto-hadronic models with the bulk of the power in hadrons.\par
Recent studies have examined the interaction between outflows and circumnuclear material in active galaxies, including the impact of cloud obscuration on emergent radiation (e.g., Mou \& Wang 2021 \cite{mou2021}). Here, we focus on a different regime: highly super-Eddington winds interacting with sparse broad-line region clouds, under conditions where mutual cloud obscuration is negligible. This approach is particularly relevant to transient super-accretors, such as those produced by tidal disruption events. The paper is organized as follows: in Section \ref{sect:interactions} we outline the model and determine the nature of the shocks created in the wind-cloud interactions in a variety of circumstances. We then calculate the electromagnetic nonthermal spectral energy distribution (SED) resulting from relativistic particle interactions with ambient fields and present the results in Section \ref{sect:sed}. The absorption of gamma rays within the wind is investigated in Section \ref{sect:absorption}. Section \ref{sect:neutrino} provides the estimate of the neutrino flux from these sources. Finally, Section \ref{sect:summary} is devoted to a brief discussion and a summary. We adopt cgs units throughout the paper.

\section{Wind-Cloud Interactions}
\label{sect:interactions}

\subsection{General Parameters}

We start by setting the size of the BLR as a function of the mass of the central black hole. \citet{kaspi2000} derived the size-luminosity relation from reverberation measurements of quasars. This relation is given by
\begin{equation}
R_{\rm BLR} = 0.03\left( \frac{\lambda L_{\lambda}(5100\textup{~\AA}
)}{10^{44}\,{\rm erg\,s^{-1}}} \right)^{0.7}\;{\rm pc},
\end{equation}
where $R_{\rm BLR}$ is the size of the BLR, and $L_{\lambda}(5100\textup{~\AA})$ is the AGN specific luminosity at wavelength $\lambda = 5100\,\textup{~\AA}$ in the rest frame. Two mass-luminosity relations are also derived, which can be expressed as size-mass relations:\par
\begin{equation}
\label{eq:radius1}
R_{\rm BLR}({\rm mean}) = 1.57 \times 10^{-4}\left(\frac{M_{\rm BH}}{10^{6}\,M_{\odot}}\right)^{1.28}\;{\rm pc},
\end{equation}
and
\begin{equation}
\label{eq:radius2}
R_{\rm BLR}({\rm rms}) = 2.73 \times 10^{-5}\left(\frac{M_{\rm BH}}{10^{6}\,M_{\odot}}\right)^{1.74}\;{\rm pc},
\end{equation}
depending on whether the continuous emission is adjusted to the mean or rms spectra of the AGNs. We use TDEs as our benchmark population. Observational demographics show that most TDEs occur around black holes with $M_{\rm BH} = 10^{5}-10^{8}\,M_{\odot}$ and are dominated by low-mass black holes, with a visible peak near $M_{\rm BH} \approx 10^{6}\,M_{\odot}$ \cite{wevers2017,wevers2019}. At the high-mass end, the observable TDE rate is strongly suppressed above a few $\times 10^{7.5}M_{\odot}$ because of direct capture by the event horizon \cite{kesden2012,yao2023}. We therefore adopt $M_{\rm BH} = 10^{6}\,M_{\odot}$ as a representative case for the scalings used in this paper. We consider the arithmetic mean of prescriptions (\ref{eq:radius1}) and (\ref{eq:radius2}), so for a black hole with $M_{\rm BH} = 10^{6}\,M_{\odot}$ we get $R_{\rm BLR} = 9.2\times10^{-5}\,{\rm pc}$.\par



We assume that the BLR clouds describe circular orbits around the central black hole and the wind is spherically symmetric and confined to a cone of solid angle $\Omega_{\rm w}$. We also assume, for simplicity, that all clouds are similar 
 with number density $n_{\rm c} = 10^{12}\,{\rm cm^{-3}}$ \citep{muller2020}. The velocity of the clouds is given by \citep{bentz2006}
\begin{equation}
v_{\rm c} \approx 7.9\times10^{3} \left(\frac{R_{\rm BLR}}{10^{-4}\, {\rm pc}} \right)^{-1/2}\left(\frac{M_{\rm BH}}{10^{6}\,M_{\odot}} \right)^{1/2} \,{\rm km\,s^{-1}}.
\end{equation}
Thus, 
 for the parameters adopted here we obtain $v_{\rm c} \approx  8\times 10^{3}\,{\rm km\,s^{-1}}$, well within the observed virial motions of BLR gas ($\approx 3\times 10^{3}-10^{4}\,{\rm km\,s^{-1}}$; see e.g.,~\cite{blandford1990,antonucci1993,laor2006,netzer2015}).\par

Although the canonical size–luminosity relation was originally derived from sub-Eddington AGN samples, some studies suggest that it can be cautiously extended to higher accretion regimes with appropriate considerations \cite{du2014, du2016}. In transient phenomena such as TDEs, observations have revealed the appearance of broad emission lines during the early stages of disk formation, indicating the presence of BLR-like gas on short timescales \cite{holoien2019, leloudas2019}. Thus, despite some remaining uncertainties, applying the size–luminosity relation in our modeling offers a plausible and physically motivated estimate of the characteristic spatial scale of the cloud population.\par

For a given size of the BLR, the radius of the clouds, $R_{\rm c}$, depends on the filling factor and the total number of clouds:
\begin{equation}
R_{\rm c} = R_{\rm BLR}\left(\frac{f_{\rm BLR}}{N_{\rm c}}\right)^{1/3},
\end{equation}
or, expressed more conveniently,
\begin{equation}
R_{\rm c} = 6.7\times10^{9} \left( \frac{R_{\rm BLR}}{10^{-4}\, {\rm pc}}\right) \left(\frac{f_{\rm BLR}}{10^{-6}} \right)^{1/3} \left(\frac{N_{\rm c}}{10^{8}} \right)^{-1/3}\;{\rm cm}.
\end{equation}

When the AGN becomes super-accreting, the standard thin disk model fails and powerful winds are launched from the region inner to the critical radius of the accretion disk. The critical radius is the distance measured from the black hole at which the radiation force equals the vertical component of the gravitational force on the surface of the disk; it can be approximated by $r_{\rm cr} \approx 4\dot{m}r_{\rm g}$, where $\dot{m}$ is the accretion rate normalized to the critical accretion rate \cite{fukue2004}. We set the wind terminal velocity to be of order the local escape speed at the launch radius, consistent with AGN Eddington winds launched where the electron scattering optical depth is $\tau_{\rm es} \approx 1$ and with observations showing that the measured outflow velocity is of order the escape speed at the inferred launch radius \cite{king2003, king2010, pounds2017}. Thus we approximate the wind velocity as the escape velocity at the critical radius $r_{\rm cr}$:
\begin{equation}
v_{\rm w} = \sqrt{\frac{2GM_{\rm BH}}{r_{\rm cr}}} \approx \frac{c}{\sqrt{2\dot{m}}},
\end{equation}
then
\begin{equation}
v_{\rm w} \approx 2.1\times10^{4} \left(\frac{\dot{M}}{10^{2}\dot{M}_{\rm cr}} \right)^{-1/2}{\rm km\,s^{-1}}.
\end{equation}

Neglecting inhomogeneities, the wind density is given by the continuity equation:
\begin{equation}
\label{eq:wind density}
n_{\rm w}(r) \approx 6.1\times10^{8} \left( \frac{\Omega_{\rm w}}{\pi\,{\rm sr}} \right)^{-1} \left(\frac{M_{\rm BH}}{10^{6}\,M_{\odot}} \right) \left(\frac{\dot{M}}{10^2\,\dot{M}_{\rm cr}}\right)\left(\frac{v_{\rm w}}{0.1\,c} \right)^{-1} \left(\frac{r}{10^{-4}\,{\rm pc}} \right)^{-2}\;{\rm cm^{-3}},
\end{equation}
where $\Omega_{\rm w}$ is the solid angle subtended by the wind. Although $\dot{M}$ and $v_{\rm w}$ are related, we keep them explicit in Equation (\ref{eq:wind density}) to clarify their separate roles in shaping the wind density profile. The number of BLR clouds within the wind is related to the total number of clouds~as
\begin{equation}
N_{\rm c}^{\rm w} = N_{\rm c} \frac{\Omega_{\rm w}}{2\pi}.
\end{equation}

\begin{figure}[H]
 \includegraphics[scale=0.6]{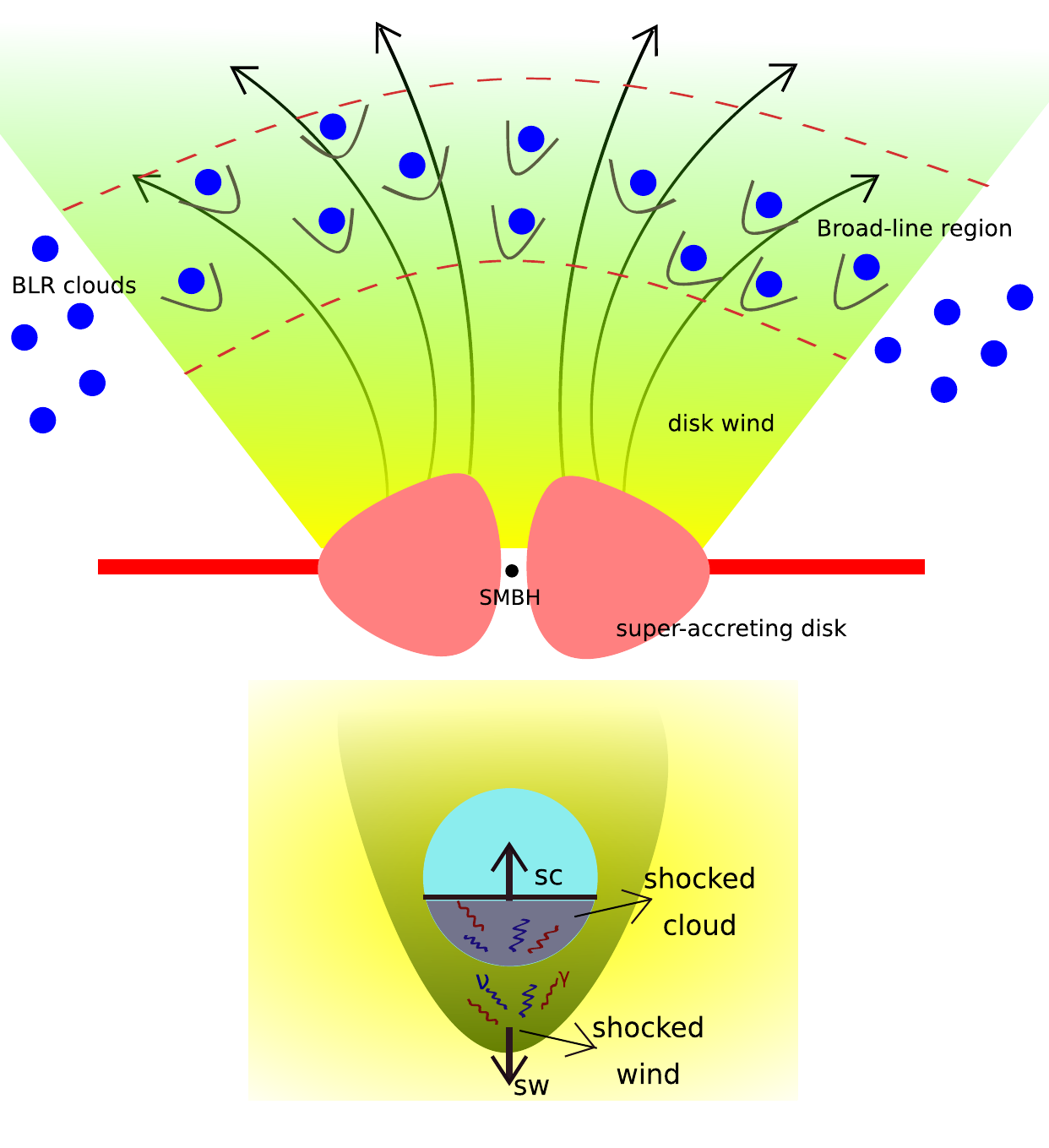}
 \caption{(\textbf{Top}): 
 Sketch of the model (not to scale). The AGN undergoes a super-Eddington accretion phase and a powerful radiation-driven wind is launched from the inner disk (labeled super-accreting disk). The outflowing supersonic wind overtakes the BLR clouds at subparsec scales from the central black hole, and bowshocks form around the clouds. (\textbf{Bottom}): Representation of a single wind-cloud interaction. In both shocked regions particles can be accelerated by DSA (Model 2), or only in one of them (Model 1). Relativistic particles interact with ambient fields and produce gamma rays (red) and neutrinos (blue).}
 \label{fig:scheme}

\end{figure}

The next step is to characterize the interactions between the BLR clouds and the wind. We present a sketch of the main elements of our model in Figure \ref{fig:scheme}. We consider two representative cases for the accretion rate: $\dot{M} = 10\,\dot{M}_{\rm cr}$ (Model 1), and $\dot{M} = 10^{5}\,\dot{M}_{\rm cr}$ (Model 2, a case of hyper-accretion that might be triggered by a extreme tidal disruption event). Since $\dot{M}_{\rm cr} = \eta\, \dot{M}_{\rm Edd}$, for a fiducial value $\eta = 0.1$, these correspond to $\dot{M} = \dot{M}_{\rm Edd}$ (Model 1), and $\dot{M} = 10^{4}\,\dot{M}_{\rm Edd}$.\par

The choice $\dot{M} = 10\, \dot{M}_{\rm cr}$ in Model 1 is observationally plausible in TDEs \cite{wevers2019, velzen2021} and transient episodes in narrow-line Seyfert 1 (NLS1) galaxies \cite{kawaguchi2003,grupe2010}. On the other hand, the extreme case $\dot{M} = 10^{5}\, \dot{M}_{\rm cr}$ is motivated by theoretical work on hyperaccretion flows in TDEs, which shows that such high accretion rates can be reached under specific conditions, such as the early fallback phase of a full stellar disruption, low radiative efficiencies, and efficient mass fallback onto the black hole \cite{coughlin2014}.\par
We note that $\dot{M}$ here refers to the accretion rate at the outer edge of the disk. In super-Eddington regimes, only a fraction $\sim \dot{M}_{\rm cr}$ is accreted onto the black hole, while the rest is ejected as a massive, optically thick wind \cite{fukue2004}. Therefore, the wind mass-loss rate $\dot{M}_{\rm w}$ is approximately equal to the input accretion rate $\dot{M}$ assumed for each model.\par
The main parameters that characterized the wind and the clouds are listed in Table \ref{table:wind-cloud}, along with the assumed values.\par

\begin{table}[H]
\caption{\label{table:wind-cloud} Parameters 
 of the wind and the clouds.}
\newcolumntype{C}{>{\centering\arraybackslash}X}
\begin{tabularx}{\textwidth}{lC}
\hline
\textbf{Parameter} & \textbf{Value}\\
\hline
Black hole mass & $M_{\rm BH} = 10^{6}\,M_{\odot}$\\
Broad-line region size & $R_{\rm BLR} = 2.9\times10^{14}\,{\rm cm}$\\
Broad-line region filling factor & $f_{\rm BLR} = 10^{-6}$ \\
Cloud radius & $R_{\rm c} = 6.1\times 10^{9}\,{\rm cm}$ \\
Cloud mass & $M_{\rm c} = 1.6\times 10^{16}\,{\rm g} $\\
Cloud density & $n_{\rm c} = 10^{12}\,{\rm cm^{-3}}$\\
Cloud velocity & $v_{\rm c} = 8\times 10^{8}\,{\rm cm\,s^{-1}}$\\
Number of clouds & $N_{\rm c}=10^8$\\
Wind solid angle & $\Omega = \pi \,{\rm sr}$\\
Average interaction radius & $r_{\rm int} = 2.6\times10^{14}\,{\rm cm}$\\
Bowshock size & $Z = 0.3\,R_{\rm c}$
\\
\hline
Model 1 & \\
Wind mass-loss rate & $\dot{M}_{\rm w} = 10\,\dot{M}_{\rm cr}$\\
Wind velocity & $v_{\rm w} = 0.22\,c$\\
Wind photosphere & $H_{\rm ph} = 1.6\times 10^{11}\,{\rm cm}$\\
\hline
Model 2 & \\
Wind mass-loss rate & $\dot{M}_{\rm w} = 10^{5}\,\dot{M}_{\rm cr}$\\
Wind velocity & $v_{\rm w} = 670.4\,{\rm km\,s^{-1}}$\\
Wind photosphere & $H_{\rm ph} = 4.6\times 10^{17}\,{\rm cm}$\\
\hline
\end{tabularx}
\end{table}

\vspace{-6pt}

\subsection{Wind Photosphere}
The wind photosphere is the surface where the optical depth $\tau_{\rm ph}$ is unity for an observer at infinity. We only consider scattering by electrons in the Thomson regime as a source of opacity, a suitable approximation for highly ionized winds. The location of the photosphere from the central black hole of the AGN, $H_{\rm ph}$, is given by \citep[][]{fukue2010}:
\begin{equation}
\tau_{\rm ph} = - \int	_{\infty} ^{H_{\rm ph}} \Gamma_{\rm w} \left(1 - \beta_{\rm w} \cos \theta \right)\kappa_{\rm co}\rho_{\rm co}{\rm d}z = 1,
\label{eq:apparent photosphere}
\end{equation}
where $\Gamma_{\rm w} = \left(1 - \beta_{\rm w}^2 \right)^{-1/2}$ is the bulk Lorentz factor of the wind, $\theta$ the viewing angle, $\kappa_{\rm co} = \sigma_{\rm T}/m_{\rm p}$ the opacity, and $\rho_{\rm co}$ is the wind density. Subscripts ${\rm co}$ refer to the quantities measured in the observer frame.\par

The photosphere is the innermost source of photons detectable from outside the AGN.  It is a thermal source with a characteristic temperature in the observer frame given by see~\cite{sotomayor2019,fukue2010}
\begin{equation}
    T_{\rm ph}^4 = \frac{L_{\rm Edd}}{4\pi H_{\rm ph}^2 \sigma_{\rm SB}\Gamma_{\rm w}^4 \left(1 - \beta_{\rm w}\cos \theta \right)^4},
\end{equation}
where $\sigma_{\rm SB}$ is the Stefan-Boltzmann constant, and we assume that the radiative power of the wind is given by the Eddington value. The photosphere plays a significant role in several nonthermal radiative processes (see Sections \ref{sect:sed} and \ref{sect:absorption} below), since it provides targets for inverse Compton scattering and $p\gamma$ collisions, as well as an absorbing field for the emerging gamma rays.\par

\subsection{Shocks}
When two fluids $1$ and $2$ collide at supersonic velocities, two shock waves are generated that propagate through each of them. In the case of two parallel, one-dimensional gas streams, \citet{lee1996} showed that the velocity of the shocks on each side of the contact discontinuity is given by (see also \cite{tenorio-tagle1981} for a detailed discussion in the context of BLR clouds colliding with the galactic disk):

\begin{equation}
    v_{\rm s1} = \frac{\gamma + 1}{2}\frac{1}{1 + \sqrt{n_{\rm 1} / n_{\rm 2}}}\left(v_{\rm 1} - v_{\rm 2}\right),
\end{equation}
\begin{equation}
    v_{\rm s2} = \frac{\gamma + 1}{2}\frac{1}{1 + \sqrt{n_{\rm 2} / n_{\rm 1}}}\left(v_{\rm 1} - v_{\rm 2}\right),
\end{equation}    
where $v_{\rm s1}$ and $v_{\rm s2}$ are in opposite directions. Assuming a monoatomic perfect gas $\gamma=5/3$ and that in the collision region $v_{\rm 1} \gg v_{\rm 2}$, we obtain:
\begin{equation}
\label{eq:veloc s1}
v_{\rm s1} = \frac{4}{3}\frac{1}{1 + \sqrt{n_{\rm 1} / n_{\rm 2}}}v_{\rm 1},
\end{equation}
and the velocity of the shock through the fluid $2$ results
\begin{equation}
\label{eq:veloc s2}
v_{\rm s2} = \frac{4}{3}\frac{1}{1 + \sqrt{n_{\rm 2} / n_{\rm 1}}}v_{\rm 1}.
\end{equation}

For the parameters of Model $1$ we obtain $v_{\rm w} = 0.22\,c$, while for Model $2$ the wind velocity is $v_{\rm w} = 670.4\,{\rm km\,s^{-1}}$. In both cases the velocity of the BLR clouds is \mbox{$v_{\rm c}~=~8\times10^{3}\,{\rm km\,s^{-1}}$}. Because of the velocity contrast in each case, we calculate the shock velocities in Model $1$ in the cloud reference frame:
\begin{equation}
\label{eq:veloc sc}
v_{\rm sw} = \frac{4}{3}\frac{1}{1 + \sqrt{n_{\rm w} / n_{\rm c}}}v_{\rm w}\quad v_{\rm sc} = \frac{4}{3}\frac{1}{1 + \sqrt{n_{\rm c} / n_{\rm w}}}v_{\rm w},
\end{equation}
and for Model $2$ we consider the wind reference frame:
\begin{equation}
\label{eq:veloc sw}
v_{\rm sw} = \frac{4}{3}\frac{1}{1 + \sqrt{n_{\rm w} / n_{\rm c}}}v_{\rm c} \quad v_{\rm sc} = \frac{4}{3}\frac{1}{1 + \sqrt{n_{\rm c} / n_{\rm w}}}v_{\rm c}.
\end{equation}

The thermal cooling length scale $R_{\Lambda}$ is the parameter that determines whether a shock is radiative or adiabatic. It depends on the cooling function $\Lambda (T)$, the density of the gas, and the shock velocity~\citep{mccray1979}:
\begin{equation}
\label{eq:cooling lengthscale}
R_{\Lambda} = \frac{1.14\times10^{-29}\;(v_{\rm sh} / {\rm km\,s^{-1}})^3}{\left(n / {\rm cm^{-3}}\right) \left(\Lambda(T) / {\rm erg\,cm^{3}\,s^{-1}} \right)}\;{\rm pc}.
\end{equation}
The cooling function in the shocked region depends, in turn, on the dominant opacity mechanism and, therefore, on the gas temperature~\citep{wolfire2003}:
\begin{equation}
\label{eq:shock emissivities}
\Lambda(T) =
\begin{cases}
7\times 10^{-27} T,\quad 10^{4}\, {\rm K} \leq T \leq 10^{5} {\rm K}\\
7\times 10^{-19} T^{-0.6},\quad 10^{5}\, {\rm K} \leq T \leq 4\times 10^{7} {\rm K}\\
3\times 10^{-27}\, T^{0.5},\quad T \geq 4\times 10^{7} {\rm K}
\end{cases}
\end{equation}
where $T = 10.9 \, (v_{\rm sh}/{\rm km \,s^{-1}})^2\;{\rm K}$.\par


If $R_{\Lambda}$ is longer than the length scale of the shocked region, the shock is adiabatic, otherwise it is radiative. 

\begin{figure}[H]
\begin{adjustwidth}{-\extralength}{0cm}
\centering
\includegraphics[scale=0.5]{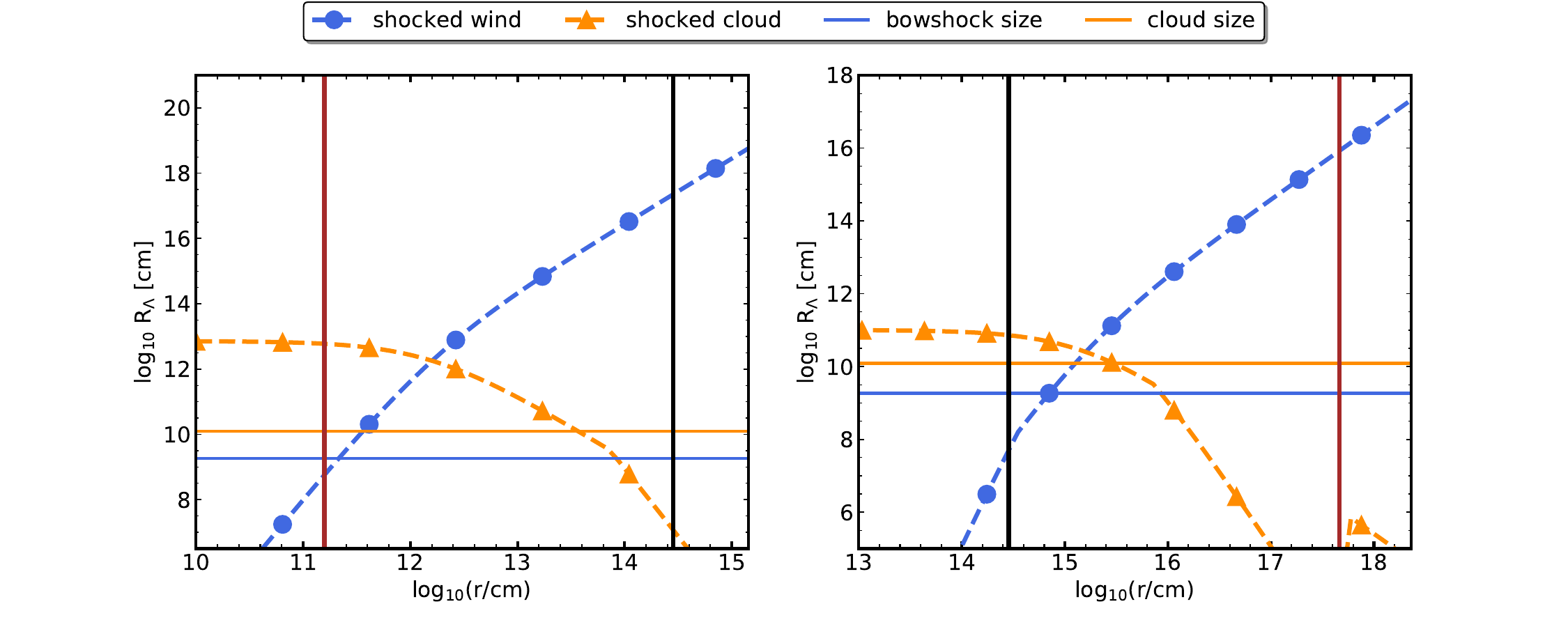}
\end{adjustwidth}
\caption{\label{fig:lengthscales} 
Length 
 scales of thermal radiative cooling for the shocks in the wind and in the cloud, along with the characteristic size scales of the shocked regions. The dashed blue and orange lines correspond to the cooling lengths of the shocked wind and shocked cloud, respectively. The horizontal lines mark the bowshock thickness (blue) and the cloud radius (orange), while the vertical lines indicate key spatial locations: the black line marks the average interaction radius $r_{\rm int} \approx 0.91\,R_{\rm BLR}$ calculated from Equation (\ref{eq:average single power}), and the brown line the wind photosphere. The shock is radiative when the cooling length is smaller than the corresponding characteristic size scale. (\textbf{Left}): Model 1. The interaction occurs just below the outer edge of the BLR, where the shocked wind is adiabatic and the shocked cloud is radiative. (\textbf{Right}): Model 2. The BLR lies deep inside the wind photosphere; the shocked wind is radiative and the shocked cloud becomes adiabatic due to the lower density contrast.}

\end{figure}

We suppose that the interactions occur at an average distance given by
\begin{equation}
\label{eq:average single power}
r_{\rm int} = \langle r \rangle = \frac{\int_{R_{\rm in}}^{R_{\rm BLR}} {\rm d}r \frac{{\rm d}N_{\rm c}^{\rm w}}{{\rm d}r}\,r}
{{\int_{R_{\rm in}}^{R_{\rm BLR}} {\rm d}r \frac{{\rm d}N_{\rm c}^{\rm w}}{{\rm d}r}}},
\end{equation}
where ${d\rm}N_{\rm c}^{\rm w}$ is the number of clouds in a differential volume ${\rm d}V_{\rm w} = \Omega_{\rm w}r^{2}{\rm d}r$, and we adopt $R_{\rm in} = 0.8\,R_{\rm BLR}$ \citep[][]{savic2020} for the inner radius of the BLR. Evaluating this expression numerically for a uniform distribution of clouds within the solid angle $\Omega_{\rm w}$, we obtain an average interaction radius $r_{\rm int} \approx 0.91\,R_{\rm BLR}$. This location is used consistently throughout the calculations and is shown in Figure \ref{fig:lengthscales} for reference.\par

In Figure \ref{fig:lengthscales} we show the thermal cooling length scale of the shocks in the wind and in the cloud for both models. We also include the characteristic size scales of the shocked regions (the cloud radius and bowshock thickness) in order to assess the nature of the shocks. A shock is radiative when the corresponding cooling length is shorter than the characteristic size; otherwise, it is adiabatic. In the left panel, we present the results for Model 1.  At the typical interaction radius, located just below the outer edge of the BLR, the shocked cloud cools efficiently (radiative), while the shocked wind is adiabatic. The wind photosphere lies well within the BLR, so the radiation produced in the interaction region escapes directly to the observer.\par


In the right panel, we show the analogous calculation for Model 2. In this case, the BLR lies well within the wind photosphere, and the situation is reversed: the shocked wind is radiative, while the shocked cloud becomes adiabatic. This behavior results from the higher density and lower velocity of the wind in this model. In Table \ref{table:scales} we show the values of the thermal cooling length scales calculated for the shocks in each model ($R_{\Lambda, \rm sw}$ and $R_{\Lambda, \rm sc}$), and the characteristic length scales ($L_{\rm w}$ and $L_{\rm c}$). Note that $t_{\rm c}$ is equal in both models because the cloud radius and velocity are assumed to be the same.\par

\begin{table}[H]
\caption{\label{table:scales}Length 
 scales and limits on timescales related to the wind and the overtaken clouds. Here $t_{\rm c}$ is the time that it takes to a cloud to enter into the wind, while $t_{\rm RT}$ and $t_{\rm KH}$ denote timescales associated with Rayleigh-Taylor and Kelvin-Helmholtz instabilities, respectively.}
\centering
\begin{adjustwidth}{-\extralength}{0cm}
\centering 
\newcolumntype{C}{>{\centering\arraybackslash}X}
\begin{tabularx}{\fulllength}{Lccccccc}
\hline
& \boldmath{$R_{\Lambda,{\rm sw}}\,[{\rm cm}]$} & \boldmath{$R_{\Lambda,{\rm sc}}\,[{\rm cm}]$} & \boldmath{$L_{\rm w}\,[{\rm cm}]$} & \boldmath{$L_{\rm c}\,[{\rm cm}]$} &  \boldmath{$t_{\rm c}\,[{\rm s}]$}  & \boldmath{$t_{\rm RT}\,[{\rm s}]$} & \boldmath{$t_{\rm KH}\,[{\rm s}]$}  \\
\hline
Model 1 & $1.9\times 10^{17}$ & $1.9\times 10^{7}$ & $1.8\times 10^{9}$ & $1.2\times 10^{10}$ & $15.4$ & $>243.2$ & $>2506.5$ \\
Model 2 & $3.3\times 10^{7}$ & $7.5\times 10^{10}$ & $1.8\times 10^{9}$ & $1.2\times 10^{10}$ & $15.4$ & $>24.3$ & $>98.5$ \\
\hline
\end{tabularx}
\end{adjustwidth}
\end{table}

Clouds can be destroyed by hydrodynamic instabilities or by effects of the thermal energy injected by the shocks. The exact calculation of the cloud  lifetime is very complex, since different processes operate simultaneously and the problem becomes highly non-linear. However, we can obtain some estimates by computing the analytical timescales~\cite{araudo2010}. Clouds, on the other hand, can be stabilized by several mechanisms: magnetic fields, radiative losses, morphology, among other properties can affect or even suppress the instabilities~\cite{cooper2009, banda-barragan2019}. Therefore, the analytical estimates of the instabilities must be considered as a lower limit. These estimates of timescales are presented in Table \ref{table:scales} for Rayleigh Taylor, $t_{\rm RT}$ and Kelvin-Helmholtz, $t_{\rm KH}$, instabilities, along with the time it takes for a cloud to enter into the wind. In any case, clouds are destroyed and formed or enter in the wind continuously. It is important to note that the time required for a cloud to enter the wind is always less than the time required for it to be destroyed. So we assumed that on average their number is roughly constant, although some level of flickering ($\sim$10\% on the shorter timescales) should be expected (see Ref. \cite{sotomayor2022} for details). This hypothesis is supported by several cloud formation models, in which clouds are launched by line-driven radiation forces from the outer parts of the disk and injected into the windy inner zone \cite{czerny2011,muller2022}.\par

\subsection{Magnetic Field in the Clouds}

The magnetic field in the clouds of the BLR of Seyfert 1 galaxies is not known. In the case of galactic molecular clouds, Zeeman splitting of emission spectral lines can be used to estimate the field strength \cite{crutcher1999}. This allows to investigate the relation of the magnetic field with the density of the clouds \cite{crutcher2010}. These results can be complemented with plane-of-sky field strengths determinations from the dispersion of polarization directions \cite{myers2021}. The maximum strength of the magnetic field seems to be fairly constant at $\sim 10$ $\mu$G up to densities $n\sim 300$ cm$^{-3}$. At densities $n > 300$ cm$^{-3}$, the field strength increases with density with a power-law exponent $\kappa\sim 2/3$ \cite{crutcher2012}. This result, however, is inconclusive, because of the uncertainties of the existing measurements. Simulations show that the actual $B-n$ relationship is strongly dependent on a number of factors, including initial conditions, level of turbulence, and geometry of the cloud \cite{Li2015}. If we extrapolate the relation $B=B_0(n/n_{0})^{\kappa}$, with $B_0=10^{-5}$ G, $n_0=300$ cm$^{-3}$, and $\kappa=0.65$, to the densities of BLR clouds, \mbox{$n_c\sim 10^{12}$ cm$^{-3}$}, we get $B\sim 15$ G. The differences, however, between the physical conditions of both types of clouds are important. At temperatures much higher than those of molecular clouds in star forming regions, there is no reason to assume the validity of the $B-n$ relationship.\par

Unlike the molecular clouds in the Galaxy, the atomic clouds in the BLR of AGNs are photoionized by the central continuum source and have temperatures of $T_{\rm c}\sim 10^4$ K. It was noted long ago by Rees \cite{rees1987} that the intercloud medium could not confine such clouds and magnetic confinement might be necessary. So, the stability of the clouds seems to require large magnetic fields. These field might be transported by the superwind launched by the disk. If the toroidal component is the dominant one, as expected, at distances of hundreds to thousands of gravitational radii \cite{spruit2010}, fields of the order of several G should pervade the intercloud~medium.\par 

The requirement of magnetic confinement imposes:
\begin{equation}
    \frac{B^2}{8 \pi}\geq 3 n_c k T_{\rm c}.
\end{equation}

In convenient units:
\begin{equation}
    B\geq \left(\frac{n_c}{10^{10}\;\rm cm^{-3}}\right)^{1/2} \left(\frac{T_{\rm c}}{10^4\; \rm K}\right)^{1/2}\;\;\; \textrm{G}.
\end{equation}
We get then fields $\sim 10$ G for clouds with densities of $n_c\sim10^{12}$ cm$^{-3}$. This value might be enhanced by shock compression up to a factor of 40 G. The magnetic energy density within the cloud, besides, should be smaller than the shock kinetic energy density, otherwise the clouds would be mechanically incompressible. This latter restriction can be written as:
\begin{equation}
    \frac{B^2}{8 \pi}= \beta \frac{9}{8} m_p n_c v_{\rm sh}^2,
\end{equation}
where $\beta<<1$ is the sub-equipartition parameter. For $\beta\sim 10^{-3}$, $n_c\sim 10^{12}$  cm$^{-3}$, and $v_{\rm sh}\sim 10^3$ km s$^{-1}$, we get $B_{\rm c}\sim 2$ G.\par

In what follows, we shall adopt the $B_{\rm sw}=40$ G for the shocked wind and \mbox{$B_{\rm c}=1$ G} for the clouds, in accordance with the considerations given above. These numbers are significantly lower than values assumed by other authors in recent papers \cite{delpalacio2019, muller2020, muller2022}, but in agreement with those values inferred from polarimetric observations \cite{silantev2013, piotrovich2017}. They can be considered as a conservative estimate.\par

\section{Nonthermal Radiation}
\label{sect:sed} 
As shown above, the regions suitable for diffusive acceleration of particles by the Fermi mechanism are different in each model: in Model 1 particles are accelerated in the shocked wind, while in Model 2 particle acceleration occurs only in the clouds. The time this process operates is limited by the lifetime of the cloud and for the duration of the super-critical accretion phase.\par 

The power injected in relativistic particles $L_{\rm rel}$ is a fraction $q_{\rm rel}$ of the kinetic power of the shock $L_{\rm k,s}$. For the shock propagating in the wind and a single wind-cloud interaction:
\begin{equation}
L_{\rm k, sw}^{1} = \frac{1}{2}n_{\rm w}\mu m_{p}v_{\rm sw}^{3}A_{\rm bs},
\end{equation}
where $A_{\rm bs}$ is the is the surface area of the bowshock. For the shock in the cloud:
\begin{equation}
L_{\rm k, sc}^{1} = \frac{1}{2} \frac{M_{\rm c}v_{\rm sc}^2}{t_{\rm cross}},
\end{equation}
where $t_{\rm cross}$ is the time the shock takes to cross the cloud.

The total kinetic power available to accelerate particles by DSA then is
$L_{\rm rel} = q_{\rm rel}L_{\rm k}$.  For Model 1 we get
\begin{equation}
L_{\rm rel} = q_{\rm rel} N_{\rm c}^{\rm w}L_{\rm k, sw}^{1},
\end{equation}
while for Model 2
\begin{equation}
L_{\rm rel} = q_{\rm rel} N_{\rm c}^{\rm w} L_{\rm k, sc}^{1}.
\end{equation}



The total power of relativistic particles is $L_{\rm rel} = L_{\rm p} + L_{\rm e}$, where $L_{\rm p,e}$ is the power of relativistic protons/electrons. We consider hadron-dominated models for the population of relativistic particles, so we set $L_{\rm p} = 100 L_{\rm e}$ for both models, as observed in Galactic cosmic rays see e.g.,~\citep{berezinskii1990}\par

Relativistic particles will lose energy through radiative cooling by synchrotron radiation (both electrons and protons), inverse Compton scattering (only electrons), relativistic Bremsstrahlung (only electrons), pion production in inelastic $pp$ collisions, and photohadronic inelastic $p\gamma$ collisions. The target photons for the inverse Compton scattering and $p\gamma$ interactions are photons from the wind photosphere. For the escape of particles, we consider: (i) convection by the bulk motion of the wind, and (ii) diffusion through the cloud. Detailed formulas for the timescales of particle cooling, convection, and diffusion can be found e.g. in 
Appendix A 
 of \citep{sotomayor2022} and references therein.\par

The maximum energy reached by the particles is estimated by equating the cooling timescale with the acceleration timescale. For diffusive shock acceleration e.g., \citep{aharonian2004}:
\begin{equation}
t_{\rm acc}^{-1}(E) = \eta \frac{ecB}{E},
\end{equation}
where $\eta$ is the acceleration efficiency:
\begin{equation}
\eta^{-1} = 20 \frac{D}{r_{\rm g}c}\left(\frac{c}{v_{\rm sh}}\right)^2,
\end{equation}
where $D$ is the diffusion coefficient, and $r_{\rm g} = E/eB$ is the particle gyroradius. In this work, we consider that $D$ is on the order of $10^{-3}$ times the value of Bohm. The suppression of charged particle diffusion can be caused by various factors, such as the presence of strong magnetic fields, shock formation, and the composition of the medium \cite{Hussein2014,drury1983}.\par

Once the maximum energies of the relativistic particles and the dominant cooling mechanisms have been determined, we calculate the particle distribution by solving the transport equation in phase space for electrons and protons. We assume a simple one-zone steady-state model. The particles are injected following a power law function with an exponential cutoff:
\begin{equation}
\label{eq:injection}
Q(E) = Q_{0}(E)^{-p} \exp \left(-\frac{E}{E_{\rm max}}\right)    
\end{equation}

From the spectrum of relativistic particles, we calculate the spectrum of the nonthermal radiation emitted. Below we summarize the results obtained for both models. The parameters adopted for calculating the nonthermal emission and some relevant results are presented in Table \ref{table:nonthermal parameters}.

\begin{table}[H]
\caption{\label{table:nonthermal parameters}Parameters for the populations of nonthermal particles.}
\newcolumntype{C}{>{\centering\arraybackslash}X}
\begin{tabularx}{\textwidth}{lC}
\hline
\textbf{Parameter} & \textbf{Value}\\
\hline
Adopted parameters \\
\hline
Minimum energy & $E_{\rm min}^{\rm e,p} = 2m_{\rm e,p}c^2$\\
Relativistic particles fraction &  $q_{\rm rel} = 10^{-1}$\\ 
Hadron-to-lepton energy ratio & $a = 100$\\
Injection spectral index & $p = 2.0$ \\
Diffusion coefficient & $D = 10^{-3}\,D_{\rm Bohm}$ \\
\hline
Calculated parameters \\
\hline
Model 1--- 
shocked wind\\
Electron maximum energy & $E_{\rm max}^{\rm e} = 11.05\,{\rm TeV}$\\
Proton maximum energy & $E_{\rm max}^{\rm p} = 11.05\,{\rm TeV}$\\
Magnetic field & $B_{\rm sw}= 40\,{\rm G}$ \\
Gas number density & $n_{\rm w} = 2.5\times10^{13}\,{\rm cm^{-3}}$ \\
Power in relativistic protons & $L_{\rm p} = 8.58\times 10^{36}\,{\rm erg\,s^{-1}}$ \\
Power in relativistic electrons & $L_{\rm e} = 8.58\times 10^{34}\,{\rm erg\,s^{-1}}$ \\
\hline
Model 2---shocked cloud\\
Electron maximum energy & $E_{\rm max}^{\rm e} = 1.84\,{\rm TeV}$\\
Proton maximum energy & $E_{\rm max}^{\rm p} = 1.84\,{\rm TeV}$\\
Magnetic field & $B_{\rm c}= 1\,{\rm G}$ \\
Gas number density & $n_c = 4\times10^{12}\,{\rm cm^{-3}}$ \\
Power in relativistic protons & $L_{\rm p} = 5.05\times 10^{41}\,{\rm erg\,s^{-1}}$ \\
Power in relativistic electrons & $L_{\rm e} = 5.01\times 10^{39}\,{\rm erg\,s^{-1}}$ \\
\hline
\end{tabularx}
\end{table}
\subsection{Model 1}

\begin{figure}[H]
\begin{adjustwidth}{-\extralength}{0cm}
 \centering
 \subfloat[Electrons]{\includegraphics[scale=0.46]{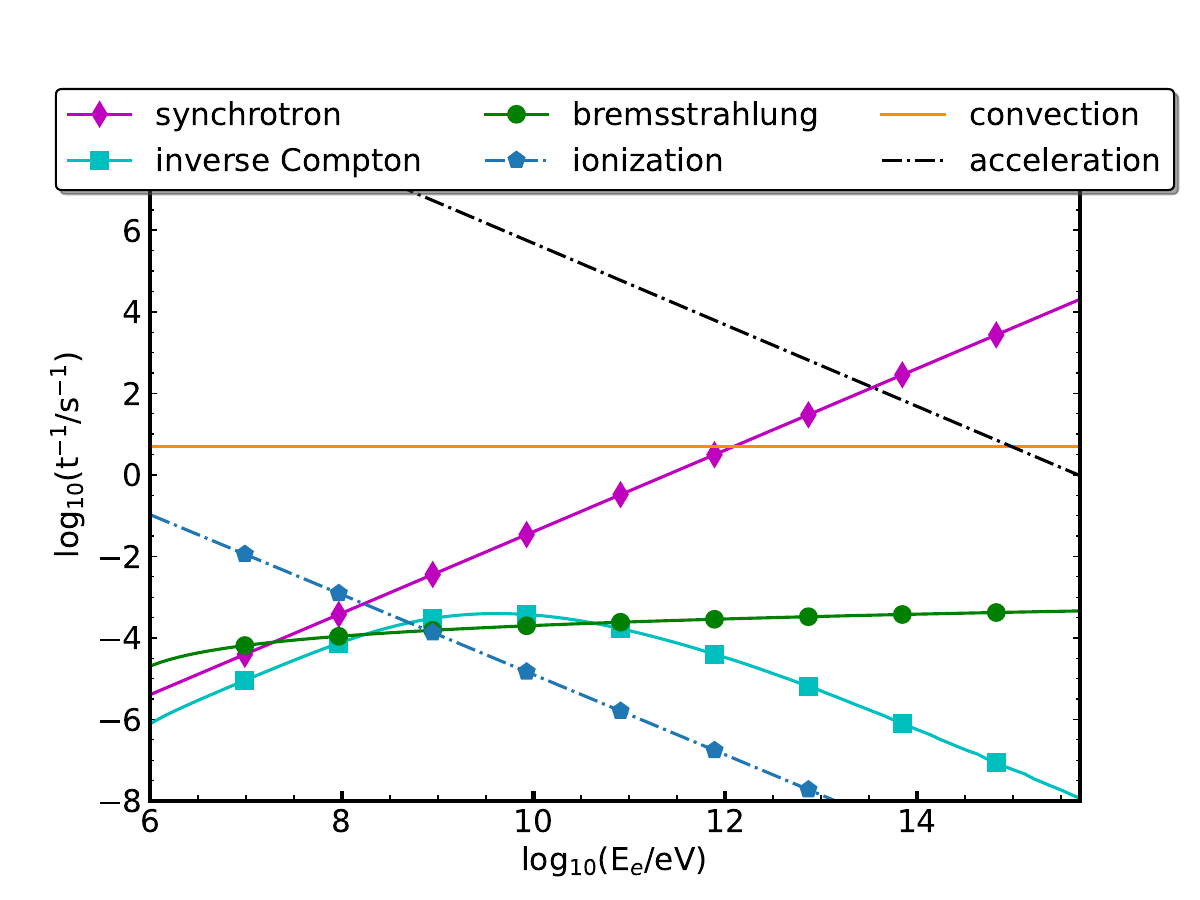}}
  \subfloat[Protons]{\includegraphics[scale=0.46]{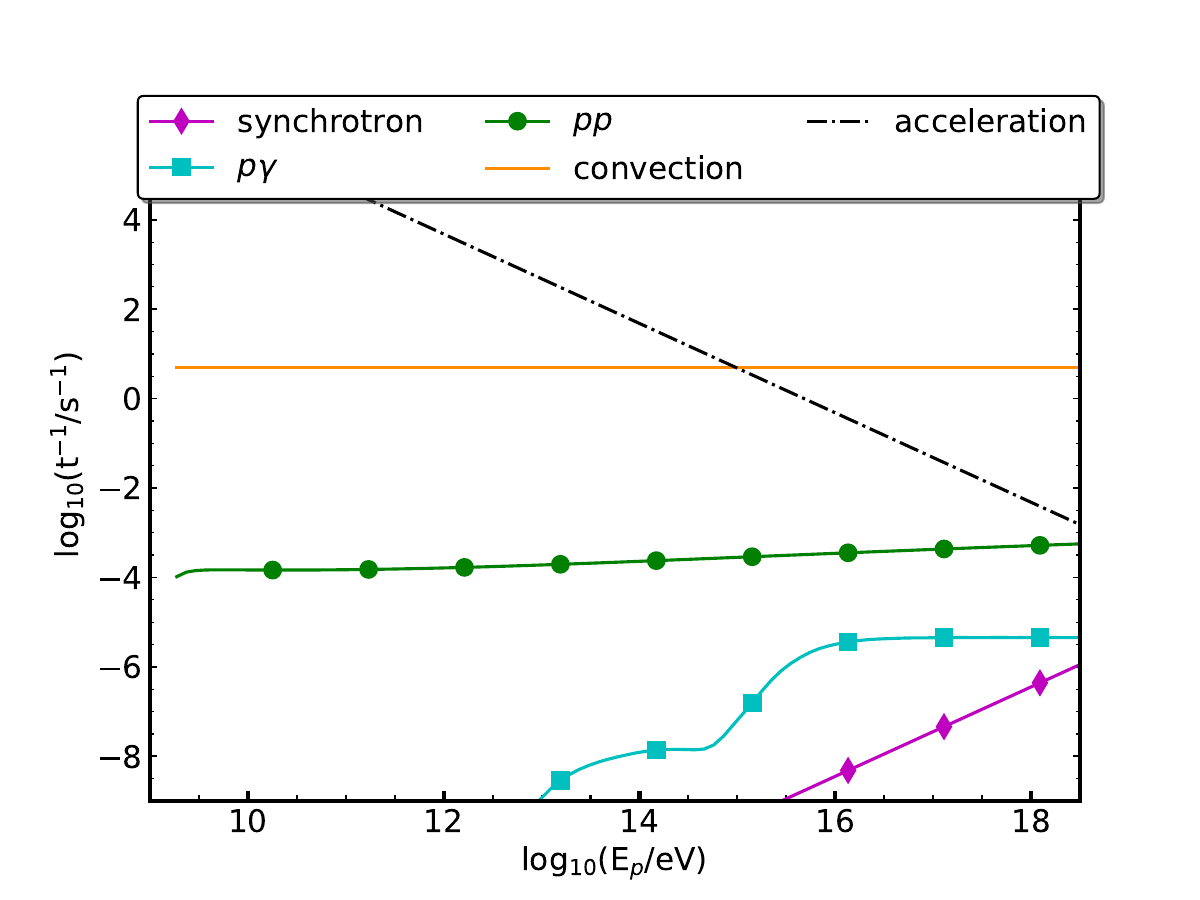}}
    \end{adjustwidth}
 \caption {Timescales 
 of acceleration, cooling, and diffusion for relativistic electrons and protons accelerated in the shocked wind for the Model 1. (\textbf{Left}): Timescales for the electrons. (\textbf{Right}): Timescales for the protons.}
 \label{fig:timescales - mod1}

\end{figure}

In this scenario, particles are accelerated by shocks within the wind above the photosphere of the disk wind. Figure \ref{fig:timescales - mod1} illustrates the timescales for cooling, escape, and acceleration of both electron and proton species. For electrons, radiative losses become significant for energies above $1\,{\rm TeV}$, and below this threshold, convection mainly removes electrons from the acceleration zone. Although the acceleration rate of electrons is balanced with the cooling rate for 50 TeV, the size of the acceleration region limits the maximum energy that particles can reach to $E_{\rm max}^{\rm e} \sim 11\, {\rm TeV}$ (Hillas criterion). For protons, we can see in the same figure that radiative losses are not significant for any range of energies, and the acceleration rate is balanced with the escape rate because of convection for $E = 1\,{\rm PeV}$ . However, like electrons, the maximum energy is restricted by the size of the acceleration zone to $E_{\rm max}^{\rm p} \sim 11\,{\rm TeV}$.\par

\begin{figure}[H]
\includegraphics[scale=0.56]{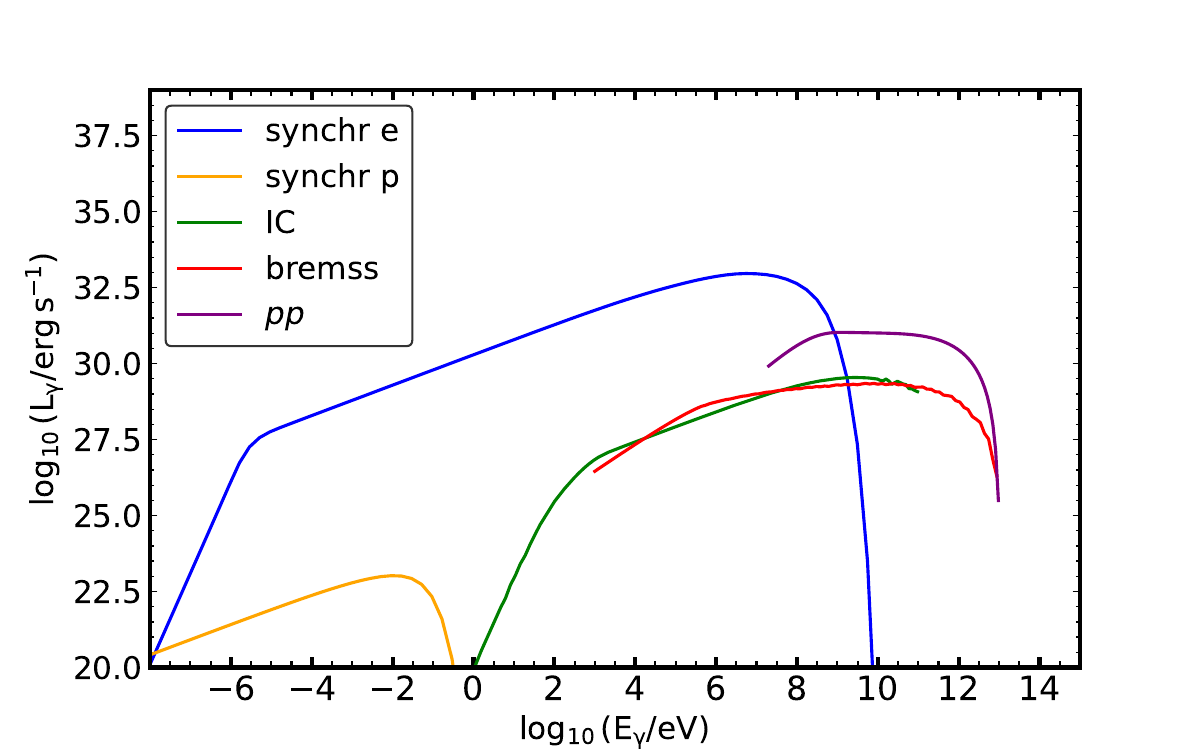}
\caption{\label{fig:sed-mod1} \small Different nonthermal contributions to the SED for Model 1. At radio energies, the SED is dominated by synchrotron radiation of electrons and protons. Gamma rays by decay of neutral pions dominate the SED at very-high energies. We assume $N_{\rm c}^{\rm w} = 5\times 10^{7}$ interactions with identical clouds at $r = 0.91\,R_{\rm BLR}$ from the supermassive black hole.}
\end{figure}

Figure \ref{fig:sed-mod1} shows the spectral energy distributions (SEDs) of the different nonthermal radiation processes taking place in the shocked wind. We show the calculation for all wind-cloud interactions. Synchrotron radiation of electrons dominates the total SED up to $E_{\gamma} \sim 100\,{\rm MeV}$ , above which gamma-ray emission from pion decay in hadronic collisions dominates the non-thermal spectrum. The maximum luminosity of this source is at $E_{\gamma} \simeq 10\, {\rm MeV}$ and is $L_{\gamma} \simeq 4.5\times 10^{32}\,{\rm erg\,s^{-1}}$, which is in good agreement with the total power available in relativistic particles of $L_{\rm rel} = 8.7\times 10^{35}\,{\rm erg\,s^{-1}}$.\par


\vspace{-6pt}

\subsection{Model 2}

\begin{figure}[H]
\begin{adjustwidth}{-\extralength}{0cm}
 \centering
  \subfloat[\small Electrons]{
    \includegraphics[scale=0.45]{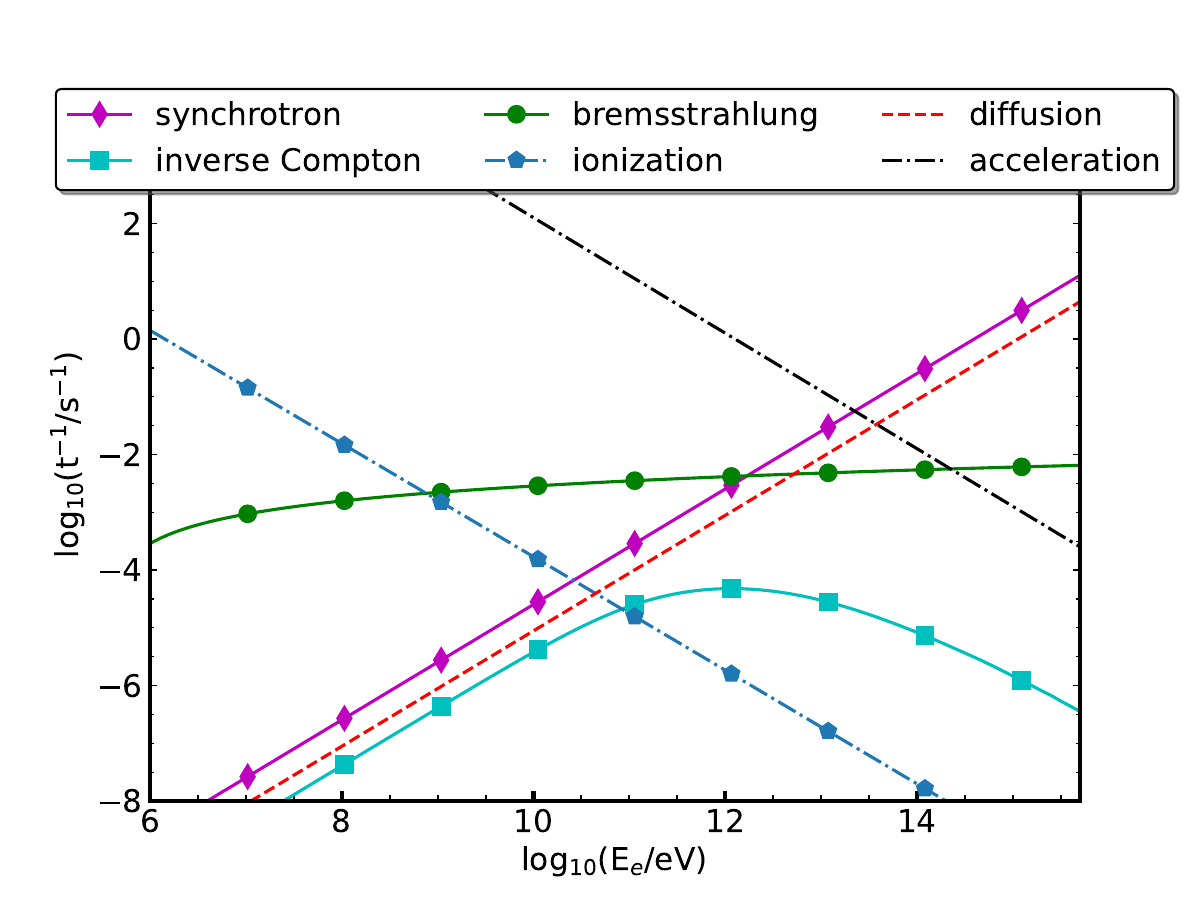}}
  \subfloat[\small Protons]{
    \includegraphics[scale=0.45]{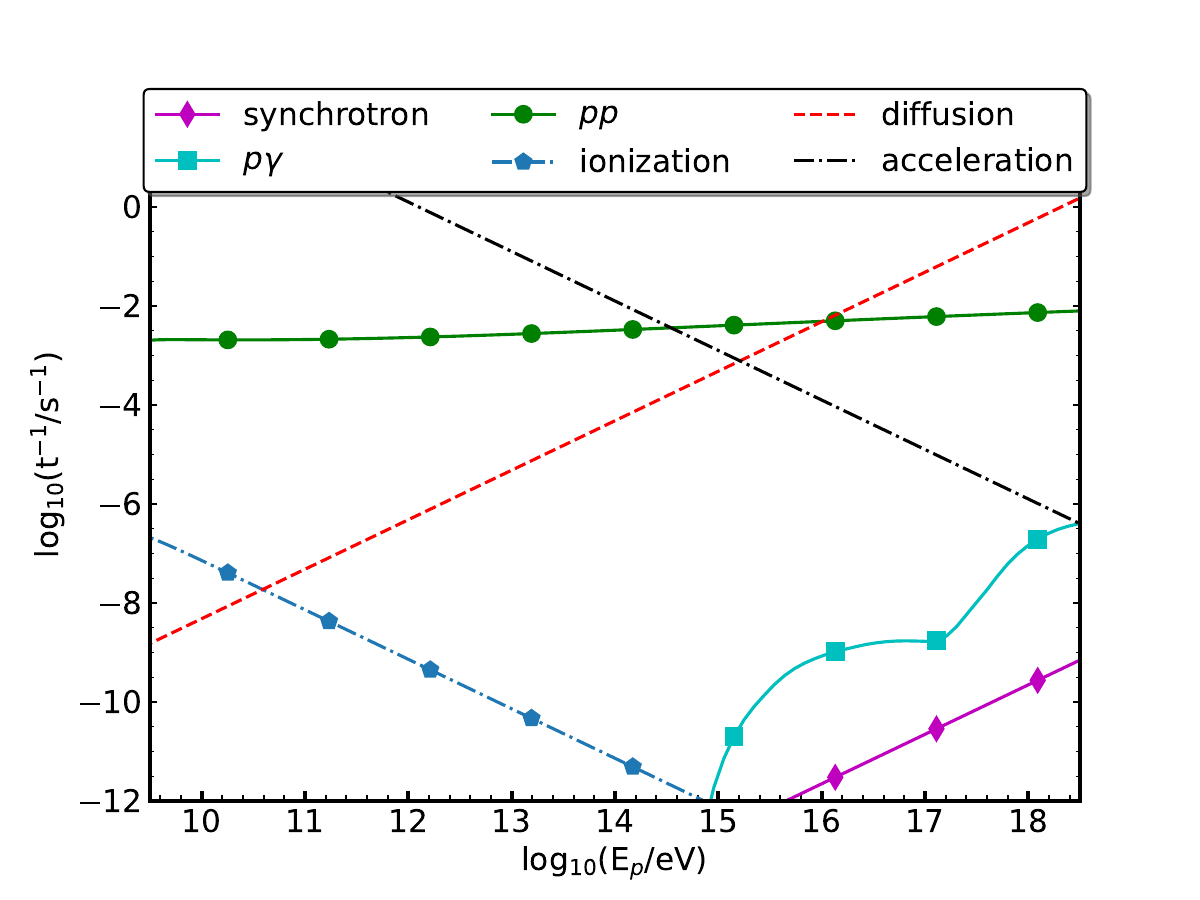}}\end{adjustwidth}
 \caption{Timescales of acceleration, cooling, and diffusion for relativistic electrons and protons accelerated in the cloud for Model 2. (\textbf{Left}): Timescales for the electrons. (\textbf{Right}): Timescales for the~protons.}
 \label{fig:timescales - mod2}

\end{figure}

Shocks within the clouds accelerate relativistic particles, which lose energy by emitting radiation. Diffusion becomes the primary escape mechanism for these particles. Figure~\ref{fig:timescales - mod2} presents the results of our calculations for all temporal scales involved. For electrons, ionization losses are relevant only up to $E_{\rm e} \sim 1\, {\rm GeV}$, beyond which Bremsstrahlung losses are the dominant cooling mechanism up to energies of $\sim 1\, {\rm TeV}$. The acceleration rate is balanced with the synchrotron cooling rate for $E_{\rm e} \sim 40\,{\rm TeV}$. However, electrons do not reach this energy as the Hillas criterion imposes a maximum energy of $E_{\rm max}^{\rm e} \sim 2\,{\rm TeV}$. The high density in the clouds makes inelastic $pp$ collisions the main cooling mechanism for protons. The cooling rate through this radiative process balances the acceleration rate for $E_{\rm p} \sim 200\,{\rm TeV}$. The Hillas criterion imposes $E_{\rm max}^{\rm p} \sim 2\,{\rm TeV}$, just as it does for electrons.\par 

\vspace{-15pt}

\begin{figure}[h!]
\includegraphics[scale=0.6]{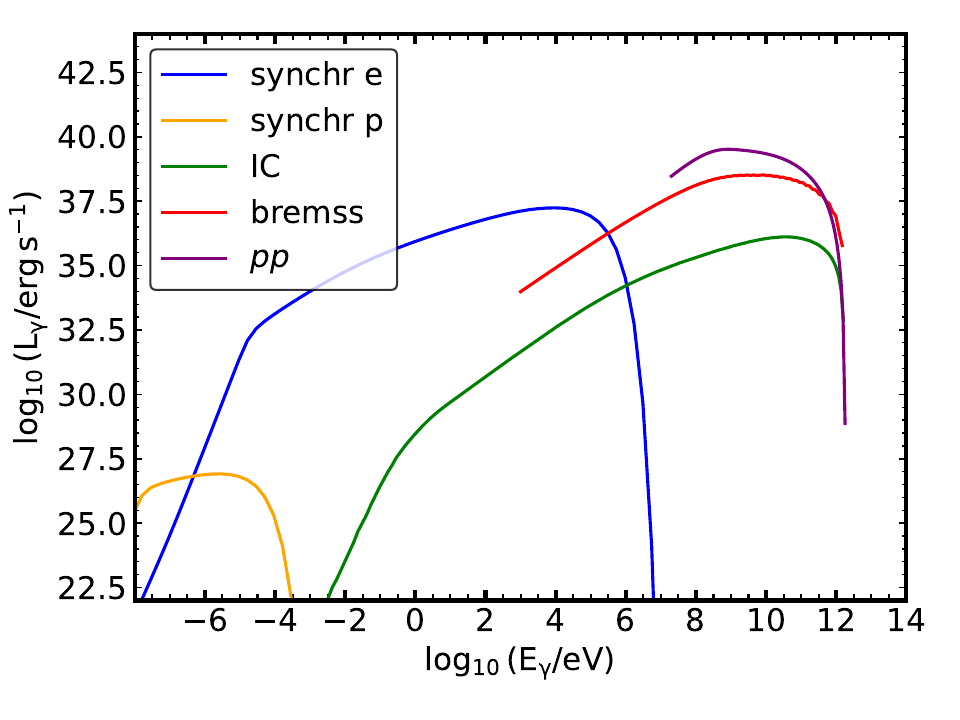}
\caption{\label{fig:sed-mod2} The same as in Figure \ref{fig:sed-mod1}, but for Model 2.}
\end{figure}

The nonthermal SEDs calculated with this model are presented in Figure \ref{fig:sed-mod2}. As in Figure \ref{fig:sed-mod1}, we present the calculation for all wind-cloud interactions. From very long wavelength radio frequencies, synchrotron radiation from electrons is the dominant emission mechanism in the total SED, with a peak luminosity of $L_{\gamma} \sim 3\times10^{37}\,{\rm erg\,s^{-1}}$ at $E_{\gamma} \sim 10\,{\rm keV}$. Soft gamma rays are mainly generated through relativistic Bremsstrahlung emission. This process is important in this model because of the high density of the target matter field. For high-energy gamma rays, the bulk of the emission is by decay of neutral pions in $pp$ collisions. The peak luminosity of the source is reached at $L \sim 10^{40}\,{\rm erg\,s^{-1}}$, at $E_{\gamma} \sim 200\,{\rm MeV}$. The high-energy cutoffs observed in the Bremsstrahlung and $pp$ components in \mbox{Figures \ref{fig:sed-mod1} and \ref{fig:sed-mod2}} arise from the exponential term in the particle injection function (Equation \eqref{eq:injection}), which limits the maximum energy attained by relativistic particles.\par

\section{Gamma-Ray Absorption}
\label{sect:absorption}

The emerging gamma-ray spectrum is attenuated by interaction with the radiation and matter fields. When clouds orbit above the wind photosphere (Model 1), the dominant gamma-ray absorption channel is photon-photon annihilation with the photosphere radiation field as the target. On the contrary, when the clouds orbit below the photosphere (Model 2), the radiation propagates in a very dense medium and the dominant absorption channel is photon interactions with the cold matter of the wind.\par

\begin{figure}[H]
\begin{adjustwidth}{-\extralength}{0cm}
\centering
  \subfloat{
    \includegraphics[scale=0.4]{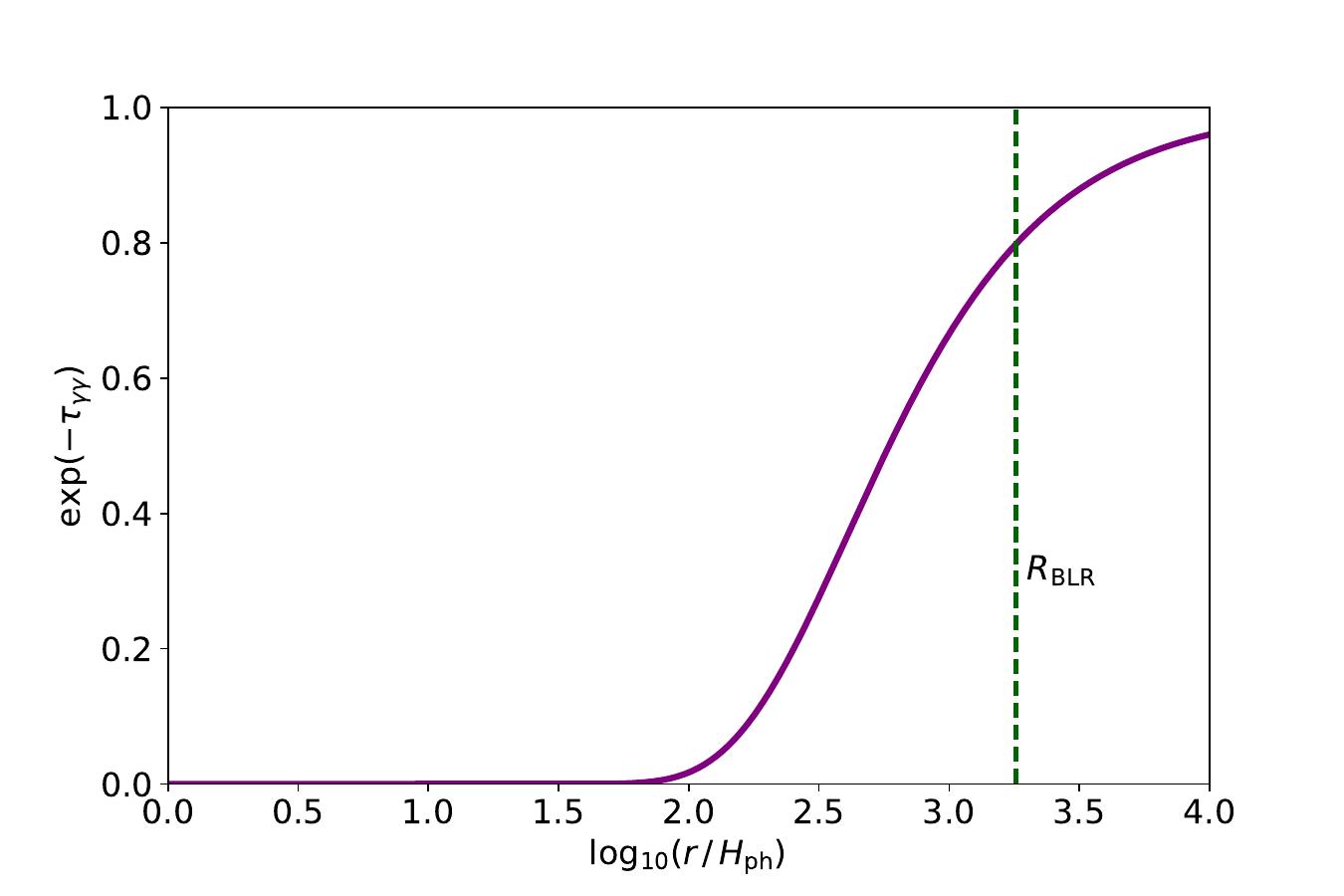}}
  \subfloat{
    \includegraphics[scale=0.38]{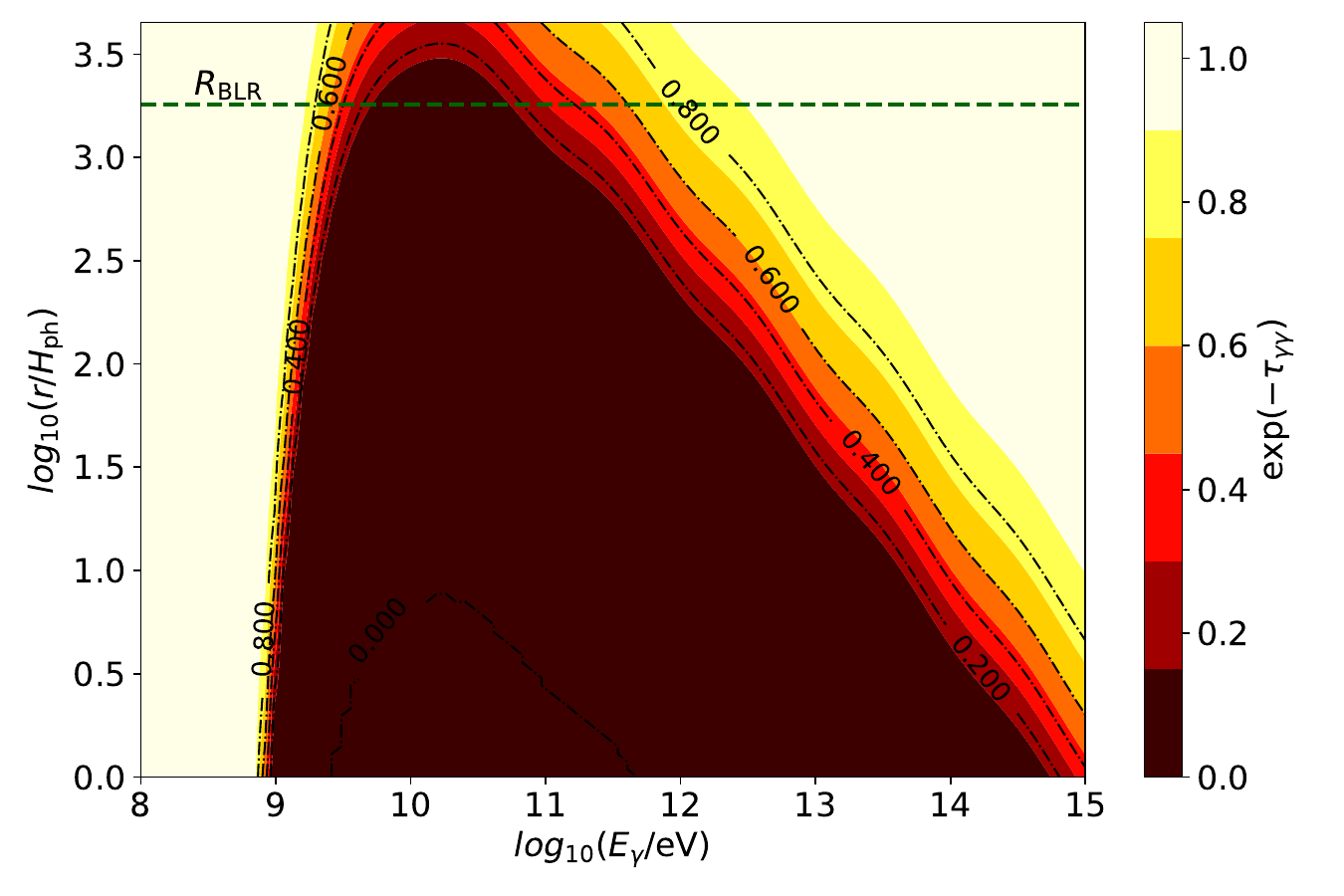}}\end{adjustwidth}
 \caption{Gamma-ray 
 attenuation by photon annihilation evaluated on the radial axis from the wind photosphere to the observer for Model 1. \emph{Left}: Attenuation calculated for a single photon with $E_{\gamma} = 1\,{\rm TeV}$. \emph{Right}: Color map for photons with energies in the range from 100 MeV to 1 PeV. The dashed green line in each plot indicates the location of the BLR.}
 \label{fig:gamma-gamma absorption}

\end{figure}

In the left panel of Figure \ref{fig:gamma-gamma absorption}, we show the optical depth for a photon with $E_{\gamma} = 1\,{\rm TeV}$ propagating in the radiation field of the photosphere in Model 1. The probability of that photon being absorbed depends on how far above the photosphere it is created. We see that when the photon is emitted at the outer edge of the BLR, the absorption probability is $P_{\rm abs} = 1 - \exp{(-\tau_{\gamma \gamma})} \simeq 0.2$, and it is absorbed for distances less than $10^{2}\,H_{\rm ph}$ from the AGN core. The calculation for photons with energies in the range between $100\,{\rm MeV}$ and $1\,{\rm PeV}$ is presented in the right panel. The radiation field of the photosphere is completely transparent for photons with $E_{\gamma} > 10\,{\rm TeV}$ or $E_{\gamma} < 100\,{\rm MeV}$ emitted at the outer edge of the BLR.\par

\vspace{-12pt}

\begin{figure}[H]
\includegraphics[scale=0.6]{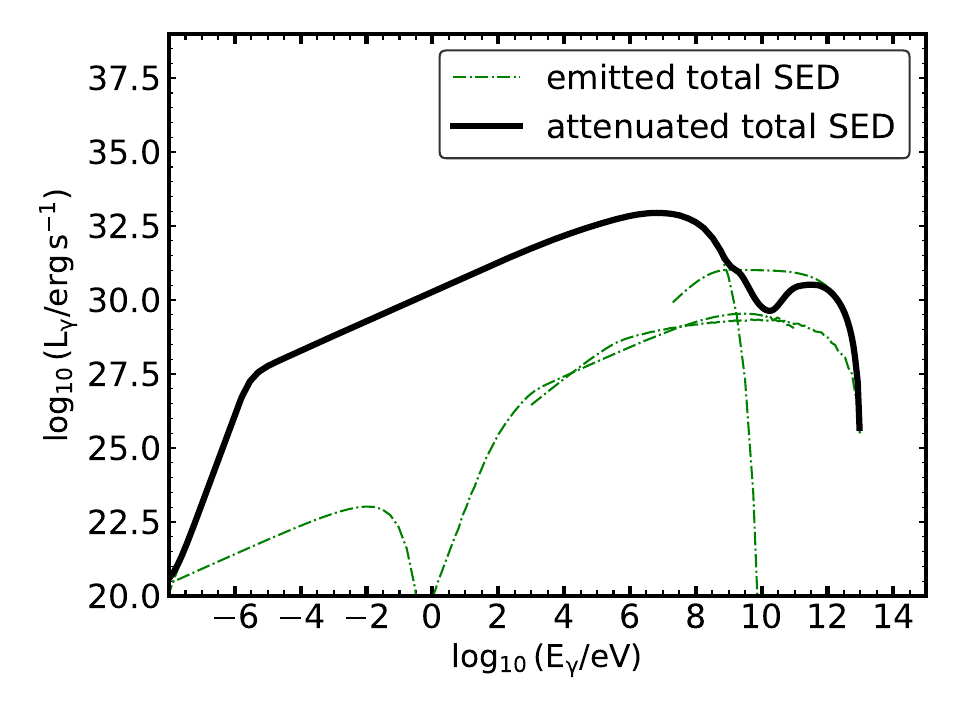}
\caption{\label{fig:corr sed-mod1} The 
 same as in Figure \ref{fig:sed-mod1}, but corrected by absorption effects.}
\end{figure}

Figure \ref{fig:corr sed-mod1} shows the nonthermal SED of Model 1 corrected by photon absorption. We see that only for energies $1\,{\rm GeV}<E_{\gamma}<1\,{\rm TeV}$ the radiation is partially attenuated. At $E_{\gamma} \sim 10\,{\rm GeV}$ there is a suppression of about two orders of magnitude of the emitted~luminosity.\par

\begin{figure}[H]
\includegraphics[scale=0.6]{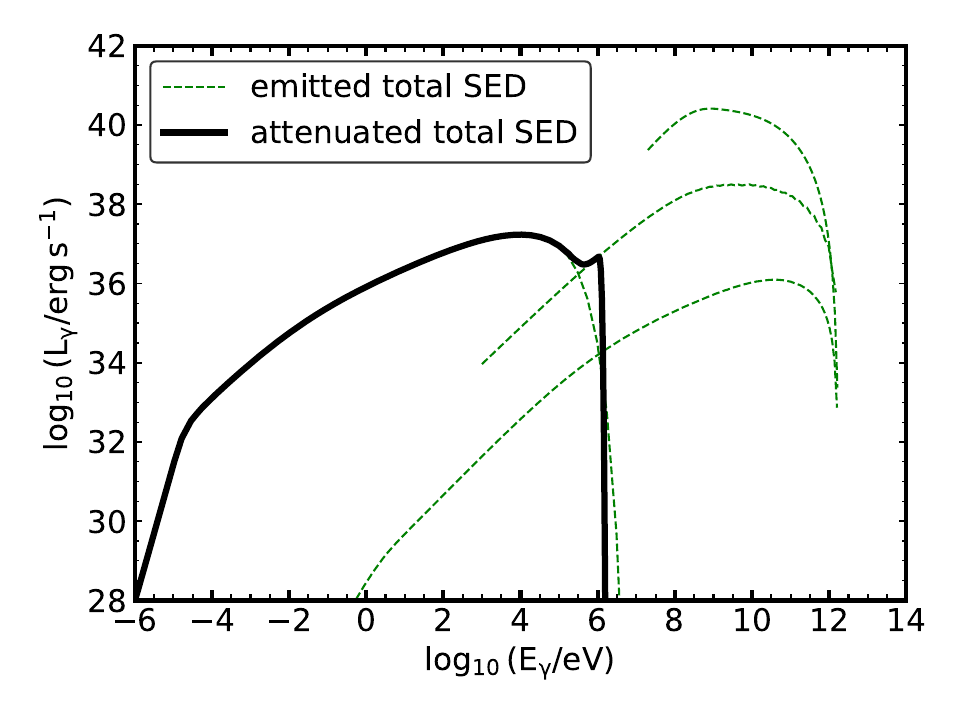}
\caption{\label{fig:corr sed-mod2} The same as in Figure \ref{fig:sed-mod2}, but corrected by absorption effects.}
\end{figure}

For Model 2 we calculate the absorption of gamma rays into the wind by $\gamma p$ collisions via pair and pion creation, and $\gamma \gamma$ annihilation with radiation from the photosphere. We obtain that all gamma rays are completely absorbed within the wind. The corrected SED is shown in Figure \ref{fig:corr sed-mod2}. Although the gamma rays are absorbed into the wind, neutrinos can escape. This scenario yields thus a source of orphan neutrinos.

\vspace{-9pt}

\section{Neutrino Production}
\label{sect:neutrino}
In inelastic $pp$ collisions, neutrinos are produced by decay of muons and charged~pions:
\begin{equation}
\pi^{-} \rightarrow \mu^{-} + \bar{\nu}_{\mu} \rightarrow {\rm e}^{-} + \nu_{\mu} + \bar{\nu}_{\rm e} + \bar{\nu}_{\mu}
\end{equation}
\begin{equation}
\pi^{+} \rightarrow \mu^{+} + {\nu}_{\mu} \rightarrow {\rm e}^{+} + \bar{\nu}_{\mu} + {\nu}_{\rm e} + {\nu}_{\mu}.
\end{equation}

When pions and muons decay without cooling, the neutrino spectrum (in units of ${\rm erg^{-1}\,s^{-1}\,cm^{-3}}$) can be calculated from (see \citet{kelner2006}):

\begin{equation}
\label{eq:neutrino injection}
Q_{\nu}^{\prime} (E_{\nu}^{\prime}) = nc\int_{0}^{1} \frac{{\rm d}x}{x} N_{\rm p}^{\prime}\left(\frac{E_{\nu}^{\prime}}{x} \right)F_{\nu}\left(x,\frac{E_{\nu}^{\prime}}{x} \right)\sigma_{pp}^{\rm inel}\left(\frac{E_{\nu}^{\prime}}{x} \right),
\end{equation}
where $x=E_{\nu}^{\prime}/E_{\rm p}^{\prime}$, $N_{\rm p}^{\prime}(E_{\nu}^{\prime} / x)$ is the proton distribution (in units of ${\rm erg^{-1}\,cm^{-3}}$), $\sigma_{pp}^{\rm inel}(E_{\nu}^{\prime} / x)$ is the cross section of inelastic $pp$ interactions, and the function $F_{\nu}(x,E_{\nu}^{\prime}/x )$ takes into account all neutrino flavors and is defined in \citet{kelner2006}. We denote by primed quantities those measured in the comoving frame. We calculate the timescales for pions and muons. In the range of possible energies, the radiative cooling of the secondary particles can be ignored and Equation~(\ref{eq:neutrino injection}) is suitable to calculate the neutrino spectrum. We can also neglect the contribution from photomeson production because even at the highest energies obtained in any of our models, this mechanism is about 4 orders of magnitude less efficient than the $pp$ channel (see Figures \ref{fig:timescales - mod1}b and \ref{fig:timescales - mod2}b). \par

The neutrino emissivity $j_{\nu}^{\prime}(E_{\nu}^{\prime})$ (in units of ${\rm erg\,s^{-1}\,cm^{-3}\,sr^{-1}\,erg^{-1}}$) is given by:
\begin{equation}
j_{\nu}^{\prime} (E_{\nu}^{\prime}) = \frac{1}{4\pi}E_{\nu}^{\prime}Q_{\nu}^{\prime} (E_{\nu}^{\prime}),
\end{equation}
and the neutrino flux (in units of ${\rm erg~s^{-1}~cm^{-2}~erg^{-1}}$) arriving at the Earth:
\begin{equation}
F_{\nu}(E_{\nu}) = \frac{D^{2}}{d^{2}_{\rm L}}\int_{V} j_{\nu}^{\prime} (E_{\nu}^{\prime}) {\rm d}V,
\end{equation}
where $D$ is the cosmological Doppler factor, $d_{\rm L}$ is the luminosity distance to the source, and $V$ is the source volume. The non-primed quantities are measured in the observer's frame.\par

\begin{figure}[H]
\includegraphics[scale=0.55]{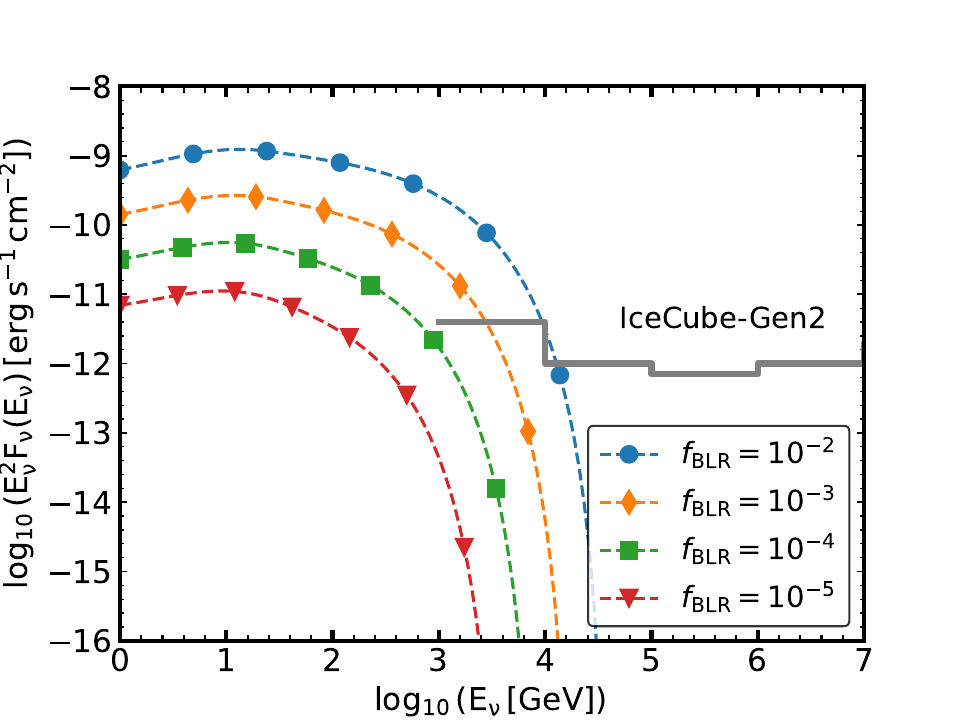}
\caption{\label{fig:neutrino1} Neutrino 
 flux for several values of the filling factor of the broad-line region in Model 2. The luminosity distance to the source is $d_{\rm L} = 5\,{\rm Mpc}$. Sensitivity of IceCube-Gen2 for a source at $\delta = 0$ with an average significance of $5\sigma$ after 10 years of observations see \citep{icecube2019}.}
\end{figure}

Figure~\ref{fig:neutrino1} shows the calculated neutrino flux for different values of the filling factor of the BLR, in the case of a nearby galaxy at $d_{\rm L} = 5\,{\rm Mpc}$. We also show the sensitivity curve of IceCube-Gen2 with an average significance of $5\sigma$ after 10 years of observations for a source at the celestial equator ($\delta=0\,{\rm deg}$) \citep{icecube2019}. For the parameters adopted in this work, the neutrino flux will only be detected by the next generation Ice Cube Observatory if the filling factor is very high: $f_{\rm BLR} = 10^{-3} - 10^{-2}$. Although high, these values are physically admissible (see \citet{osterbrock1991}).\par 

We adopt $M_{\rm BH} = 10^{6} \, M_{\odot}$ as our fiducial benchmark, motivated by typical TDE host galaxies. To demonstrate the applicability of our framework to more massive systems still consistent with both TDEs and nearby AGNs, we also consider $M_{\rm BH} = 10^{7}\,M_{\odot}$. At a fixed Eddington ratio, the neutrino power increases approximately in proportion to the black hole mass—reflecting wind energetics that scale with $L_{\rm Edd}$—while the spectral cutoffs exhibit only a modest shift toward higher energies \cite{king2015}.\par 

\begin{figure}[H]
\includegraphics[scale=0.55]{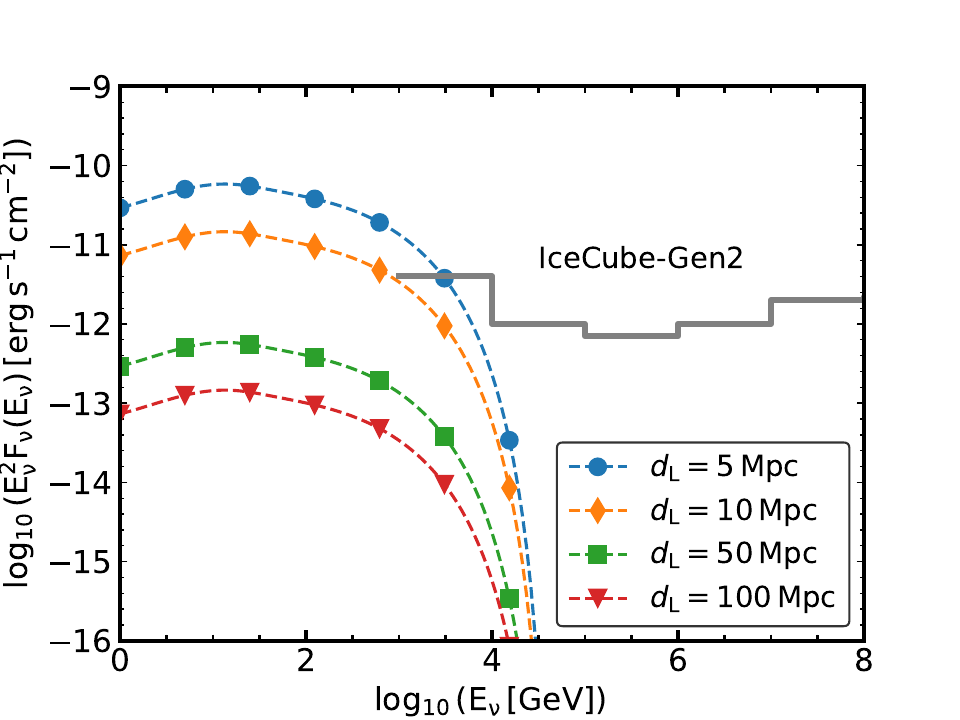}
\caption{\label{fig:neutrino2} The same as in Figure~\ref{fig:neutrino1}, but for $M_{\rm BH} = 10^{7}\,{M_{\odot}}$, $\dot{M} = 10^{5}\,\dot{M}_{\rm cr}$, and varying distance to the source. The filling factor in the broad-line region is fixed to $f_{\rm BLR} = 10^{-6}$.}
\end{figure}

\begin{figure}[H]
\includegraphics[scale=0.55]{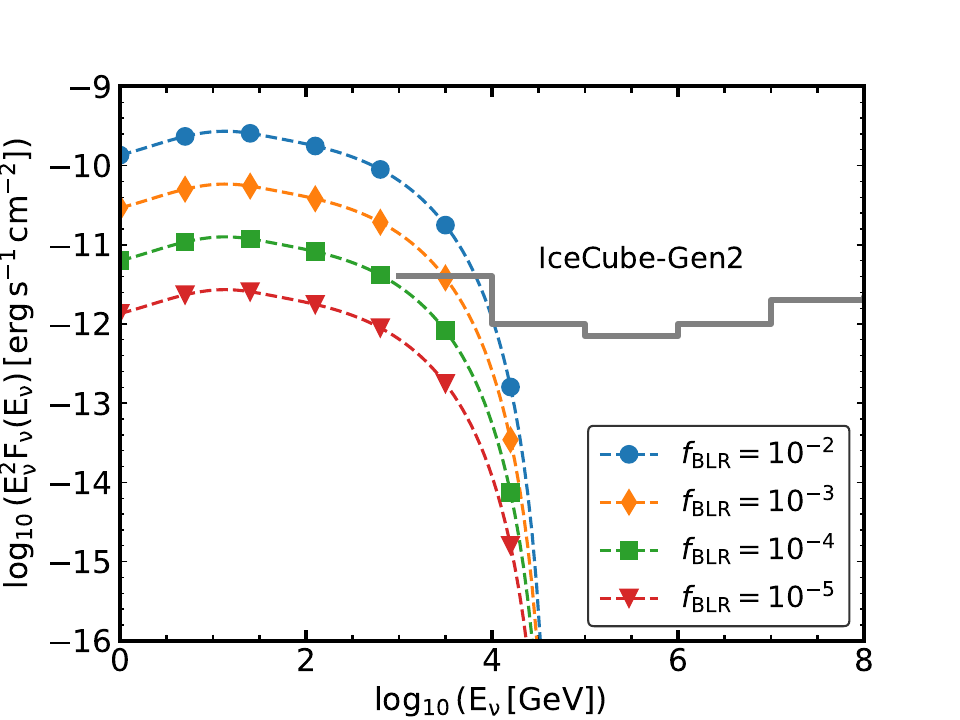}
\caption{\label{fig:neutrino3} The same as in Figure~\ref{fig:neutrino2}, but for $d_{\rm L} = 50\,{\rm Mpc}$, and varying filling factor of the broad-line~region.}
\end{figure}

For $M_{\rm BH} = 10^{7}\,M_{\odot}$, Figure \ref{fig:neutrino2} shows the flux of neutrinos arriving at the Earth for a filling factor $f_{\rm BLR} = 10^{-6}$. In this case the emission is not detectable, except for a marginal signal for very nearby galaxies at $d_{\rm L} \approx 5\,{\rm Mpc}$. Increasing the filling factor favors detection. Figure \ref{fig:neutrino3} presents the neutrino flux for an AGN at $d_{\rm L} = 50\,{\rm Mpc}$ for several values of the filling factor. The neutrino flux of an AGNs with $f_{\rm BLR} \geq 10^{-3}$ would be detectable with IceCube-Gen2 at such a distance.\par

\vspace{-9pt}

\vspace{-9pt}

\begin{figure}[H]
 \includegraphics[scale=0.55]{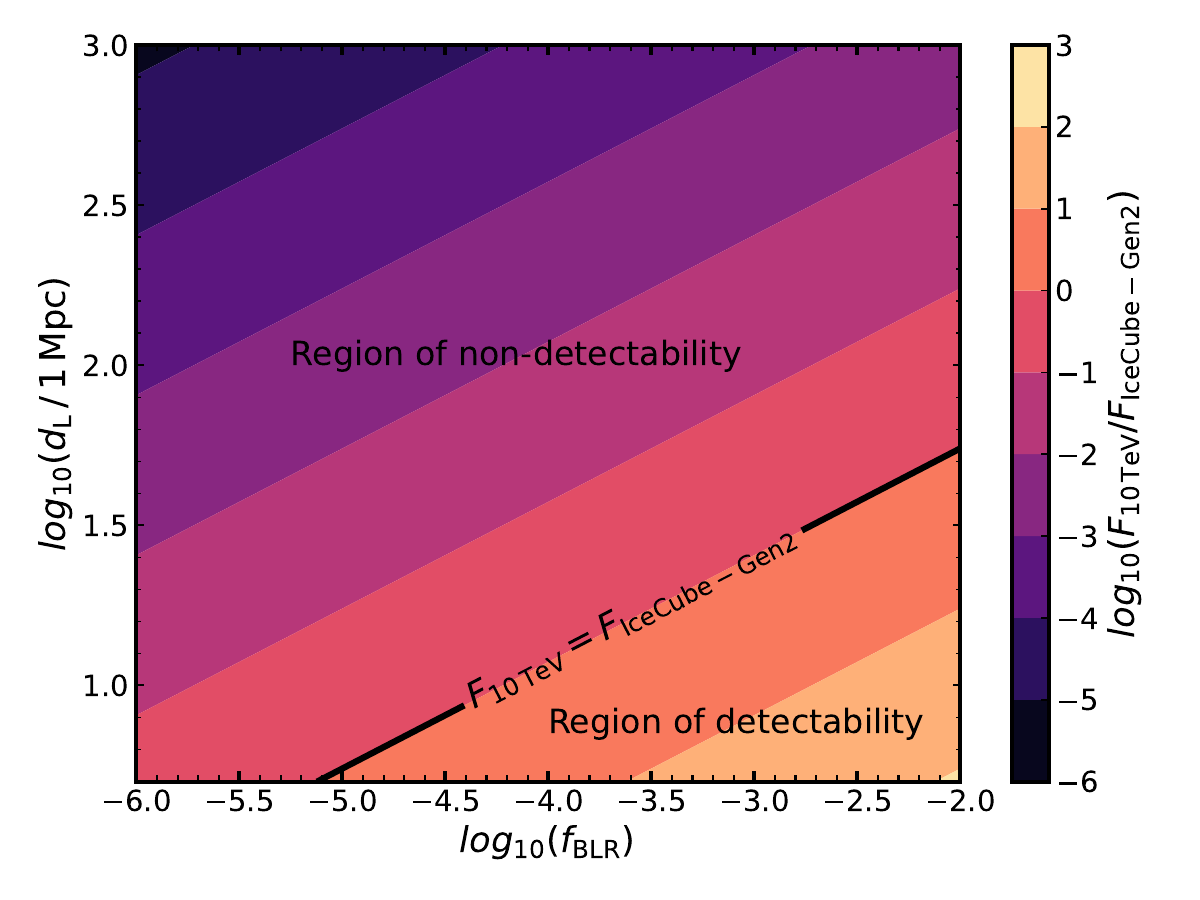}
 \caption{\small Neutrino flux at $E_{\nu}=10\,{\rm TeV}$ relative to the threshold value for IceCube-Gen2. We consider $M_{\rm BH} = 10^{7}\,M_{\odot}$, $\dot{m} = 10^{5}$, and $f_{\rm BLR}$ and $d_{\rm L}$ are in the range $(f_{\rm BLR}\times d_{\rm L}) = [10^{-5},10^{-2}]\times [5\,{\rm Mpc},1\,{\rm Gpc}]$. The thick black line indicates the limit of detectability.}
 \label{fig:regions}
\end{figure}

In Figure \ref{fig:regions} we show the detectability region in the space $f_{\rm BLR}$ vs $d_{\rm L}$ for $M_{\rm BH}=10^7$ and $\dot{m}=10^{5}$, for neutrino energy of $E_{\nu} = 10\,{\rm TeV}$. We see that for such high accretion rates, corresponding to tidal disruption events, neutrinos from Seyfert galaxies might be detectable up to distances of 30 Mpc, depending of the BLR filling factor.  



In neither case are the sources detectable by current observational instruments.\par

Although the physics of accretion and outflows in TDEs is less well constrained than in regular AGN, recent observations have revealed powerful, mildly relativistic winds launched during the early fallback phase (e.g., \cite{miller2015, alexander2016}). These outflows can carry substantial kinetic energy and may interact with the surrounding circumnuclear medium—potentially composed of residual gas from previous AGN activity, bound stellar debris, or disrupted stellar envelope material—thereby driving shocks capable of accelerating particles to relativistic energies.\par

Although it remains uncertain whether TDEs are accompanied by persistent broad-line regions, a growing body of observational evidence indicates that dense, line-emitting gas can form transiently during the flare’s evolution. Several TDEs have exhibited broad emission lines in their optical spectra, consistent with high-velocity material photoionized by the central UV/X-ray source (e.g., \cite{hung2019, nicholls2020}). Photoionization modeling suggests that such material can sustain BLR-like physical conditions over timescales of weeks to months—comparable to the expected duration of neutrino production (e.g., \cite{leloudas2019}). Furthermore, numerical simulations support the formation of compact accretion disks and extended, radiation-dominated envelopes in TDEs, both capable of launching winds and interacting with ambient gas (e.g., \cite{metzger2016, lu2020}).\par

In this context, our model offers a physically motivated framework for describing wind–cloud interactions and their associated nonthermal emission, including neutrinos. The framework is also directly applicable to NLS1 galaxies, where the presence of persistent outflows, circumnuclear clouds, and broad-line regions is well established. In NLS1s, the physical structure is comparatively better constrained, and the proposed wind–cloud interaction mechanism can operate continuously, rendering these systems particularly promising candidates for long-term neutrino emission.

\section{Final Remarks and Summary}
\label{sect:summary}
We have calculated the neutrino flux from non-blazar AGNs  undergoing a super-Eddington accretion phase. The emission originates in the interaction of the radiation-driven wind launched by the inner accretion disk with clouds of the broad line region. Shocks in the wind and shocked clouds can become, under some conditions, suitable sites of acceleration of cosmic rays and production of nonthermal radiation and neutrinos. The electromagnetic radiation might be absorbed by the wind (except at radio frequencies, see Ref. \cite{sotomayor2022}). But the neutrinos, mostly arising from $pp$ collisions in the dense medium, can freely escape. Our study focuses mainly on low-mass black holes $M_{\rm BH} \sim 10^{6}\,M_{\odot}$, which is the typical value expected for galaxies hosting TDEs, although we have investigated more massive black holes such as those usually found in NLS1 galaxies (i.e., $M_{\rm BH} \sim 10^{7}\,M_{\odot}$). Since most of these galaxies have fast clouds orbiting the central black hole and they use to go through phases of high accretion, the interaction of the wind with the clouds is~unavoidable.\par 

When the clouds are inside the photosphere, the flux of neutrinos is enhanced and the electromagnetic radiation absorbed. This creates dark neutrino sources, that only manifest through the thermal photospheric emission, typically falling in the UV region. In such cases, the neutrino flux by wind-cloud interactions could be detected by IceCube-Gen2 for galaxies with high BLR filling factors up to $\sim 100\,{\rm Mpc}$.\par 

The assumption of independent interactions between the wind and BLR clouds is justified by the low volume filling factor adopted in our models. For a black hole mass of $M_{\rm BH} = 10^6\,M_{\odot}$ with $f_{\rm BLR} = 10^{-6}$, this corresponds to mean cloud separations of $\sim 10^{12}$~cm, substantially exceeding the characteristic shock lengths generated by individual interactions (ranging from $0.3\,R_{\rm c}$ in Model 1 to $R_{\rm c}$ in Model 2, with $R_{\rm c} \sim 6\times 10^{9}$~cm). Mutual obscuration between clouds is therefore unlikely to affect our results. Gamma-ray absorption may still occur within individual clouds; however, the dominant attenuation mechanism remains the dense photon field of the wind photosphere (see \cite{sotomayor2022}). A detailed treatment of mutual obscuration effects in cloud–outflow interactions is provided by \cite{mou2021} in the context of less extreme accretion regimes. While our work focuses on super-Eddington winds and sparse cloud distributions, their analysis offers valuable insights into radiative transfer and emission signatures in denser environments.\par

The emission time is limited by the duration of the super-accretion phase of the AGN. The timescale of tidal disruption events depends on the mass of the disrupted star and the mass of the black hole. Typically, they can last several years, especially if the star is of early types. In addition, the clouds are eventually destroyed by the wind and replaced by new clouds, either entering from the sides of the wind cone or overtaken by its expansion. This should lead to some level of flickering (typically of a few percentage, see Ref. \cite{sotomayor2022}) in the continuum radio emission as well as in the total intensity of the Balmer and other relevant lines. The timescale of such flickering would depend on the specific conditions of the sources and the prevailing instabilities, but should be in the range from hours to days. Hence, observations with radio telescopes in the continuum and with optical-UV telescopes (for the lines) would provide an independent diagnostics of the candidate galaxies: sources displaying strong stochastic flickering are likely undergoing numerous wind-clouds interactions and might be associated with multi-TeV neutrino detection in the~future.\par   

Our model can be applied to other kind of obstacles that the wind might find on its way such as a population of early-type stars or neutron stars. Such objects are expected to abound in the inner regions of AGNs. Detailed calculations should be done in order to determine the impact of such populations in the neutrino emission.\par

\vspace{6pt}
\authorcontributions{Conceptualization, Gustavo E. Romero; Methodology, Gustavo E. Romero and Pablo Sotomayor; Software, Pablo Sotomayor; Formal analysis, Gustavo E. Romero and Pablo Sotomayor; Investigation, Gustavo E. Romero and Pablo Sotomayor; Writing – original draft, Gustavo E. Romero; Writing – review \& editing, Gustavo E. Romero and Pablo Sotomayor; Supervision, Gustavo E. Romero.}

\funding{This research was funded by grants grant PID2022-136828NBC41/AEI/10.13039/501100011033/ (State Agency for Research of the Spanish Ministry of Science and Innovation) and PIP 0554 (CONICET)}

\dataavailability{The datasets generated and analyzed during the current study are available from the corresponding author upon reasonable request.} 

\acknowledgments{We thank Leandro Abaroa for fruitful discussions.}
%
%

\conflictsofinterest{The authors declare no conflicts of interest.}

\begin{adjustwidth}{-\extralength}{0cm}
\printendnotes[custom] 

\reftitle{References}
\end{adjustwidth}



\begin{adjustwidth}{-\extralength}{0cm}

\PublishersNote{}
\end{adjustwidth}
\end{document}